\documentclass[sn-mathphys-num]{sn-jnl}

\hypersetup{pdfpagemode=UseNone}

\usepackage{graphicx}%
\usepackage{multirow}%
\usepackage{amsmath,amssymb,amsfonts}%
\usepackage{amsthm}%
\usepackage{mathrsfs}%
\usepackage[title]{appendix}%
\usepackage{xcolor}%
\usepackage{textcomp}%
\usepackage{booktabs}%
\usepackage{algorithm}%
\usepackage{algorithmicx}%
\usepackage{algpseudocode}%
\usepackage{listings}%

\usepackage{bm}
\usepackage{array}
\usepackage{hyperref}

\theoremstyle{thmstyleone}%
\newtheorem{theorem}{Theorem}
\newtheorem{proposition}[theorem]{Proposition}%

\theoremstyle{thmstyletwo}%

\theoremstyle{thmstylethree}%

\begin{document}

\title[A structured parametrization of the Grassmann distribution for categorical and ordinal variables]{A structured parametrization of the Grassmann distribution for categorical and ordinal variables}

\author*[1]{\fnm{Takashi} \sur{Arai}}\email{takashi-arai@sci.kj.yamagata-u.ac.jp}

\affil*[1]{\orgdiv{Faculty of Science}, \orgname{Yamagata University}, \orgaddress{\street{1-4-12 Kojirakawa-machi}, \city{Yamagata-shi}, \postcode{990-8560}, \state{Yamagata}, \country{Japan}}}

\abstract{
    We propose a multivariate probability distribution for categorical and ordinal random variables based on the Grassmann distribution.
    By imposing a specific structure on the parameter matrix of the Grassmann distribution, the proposed distribution realizes the co-occurrence probabilities required for categorical and ordinal variables.
    We also propose the variant of the diagonally dominant parametrization that is compatible with the proposed parameter structure and ensures nonnegativity of joint probabilities.
    We numerically validate the parameter estimation and the representational flexibility of the proposed model using real and artificially generated data.
    The numerical experiments demonstrate that the proposed model reproduces the correlations of dummy variables with few categories or ordinal levels reasonably well.
    We also investigate the sampling distribution of the maximum likelihood estimator and provide numerical evidence that the estimator is consistent.
    We compare the proposed model with the existing log-linear model and latent class model.
    Although the representational flexibility of the proposed model is limited compared to the existing model, the proposed model has favorable scalability in terms of computational cost and memory usage.
}

\keywords{Grassmann distribution, Ising model, P matrix}
\pacs[MSC Classification]{62H05, 62H25}

\maketitle

\newpage

\section{Introduction}
In data analysis, qualitative data such as categorical and ordinal variables often arise.
Examples include demographic attributes such as gender and occupation, questionnaire responses, the results of sensory testing measured on a Likert scale~\citep{Likert1932}, and the presence of mutations in DNA and amino acid sequences.
In today's data-rich era, these qualitative data, including nominal and ordinal variables, are abundant.
Consequently, there is a growing need to develop appropriate statistical models for analyzing qualitative data.
As a building block for such models, it is also desirable to have a probability distribution that can describe statistical association among qualitative variables.

One way to model statistical association among categorical variables is to introduce latent variables.
In latent variables models, the observed variables are assumed to be conditionally independent given the latent variables.
The statistical association among observed variables is expressed through the latent variables.
One such model that has been developed and is widely used in the field of educational measurement and psychometrics is Item Response Theory or generalized linear latent variable models~\citep{Lord1980, Reckase2009, Bock1981, Bartholomew2011}.
This model introduces continuous latent variables and adopts the categorical distribution for the conditional distribution given the latent variables.
In the fields of machine learning, the same model is referred to as exponential family principal component analysis~\citep{Murphy2012, Collins2001, Mohamed2008}.
However, the introduction of continuous latent variables often makes the marginal distribution of the observed variables analytically intractable.
This drawback complicates parameter estimation, necessitating the use of a computationally demanding and time-consuming approximation technique.
Another popular model for the multivariate categorical variables is the latent class model~\citep{Agresti2012}.
This model introduces discrete latent variables and is essentially equivalent to the mixture model of categorical distributions.
However, in the latent class model, the number of latent classes required to accurately reproduce associations among observed variables can increase rapidly as the number of observed variables increases.
Moreover, maximum likelihood estimation in latent variable models can suffer from numerical instability, local optima, and improper solutions, which causes a problem in interpreting the estimated latent variables.

Another way to model associations among categorical variables is to use log-linear models~\citep{Agresti2012}.
This model is an extension of the Ising model~\citep{Ising1925} to categorical variables and relates the logarithm of joint probabilities to the model's interaction parameters.
Commonly, models that have up to the second-order interactions are adopted for the log-linear model.
In this case, the model represents joint probabilities using second-order moments of the categorical variables.
The conditional distribution of the log-linear model is also a log-linear model.
Furthermore, the log-linear model with interactions up to second order can be viewed as the maximum-entropy distribution among distributions that reproduce the first- and second-order moments of the dummy variables.
In this sense, the log-linear model is a natural distribution for multivariate categorical variables.
However, as in the case of the Ising model, the model has a difficulty of the computational complexity in calculating the normalization constant as the number of the categorical variables increases.

Recently, a multivariate distribution for correlated binary variables based on a Grassmann number has been proposed~\citep{Arai2021}.
Hereafter, we refer to this distribution as the Grassmann distribution.
Also, a probability distribution that describes correlation between binary and continuous variables has been developed~\citep{Arai2022}.
The Grassmann distribution has similar properties to the multivariate normal distribution.
Specifically, the marginal and conditional distribution of the Grassmann distribution is also a Grassmann distribution.
These properties offer computational advantages in estimating model parameters.
Furthermore, there is no need to sum over all possible states explicitly when computing the normalization constant of the distribution.
The Grassmann distribution is similar in spirit to the log-linear model with interactions up to second order, which represents joint probabilities using the first and second moments of dummy variables.
However, a naive application of the Grassmann distribution to categorical and ordinal dummy variables fails to find the optimal solution for parameter estimation; it may lead to numerically pathological solutions for categorical variables and often converge to a local optimum rather than the global optimum for ordinal variables.
These problems may stem from the fact that certain co-occurrence probabilities of categorical and ordinal dummy variables are exactly zero.
In such cases, it may be difficult for the Grassmann distribution to numerically find the parameter structure of the Grassmann distribution that realizes the co-occurrence patterns.
Therefore, in this paper, we propose a parameter structure of the Grassmann distribution that realizes the co-occurrence patterns of categorical and ordinal dummy variables.

This paper is organized as follows.
In Sec.~\ref{sec:result}, we  propose the structured parametrization and state the main results as propositions with proofs.
We also clarify that this structured parametrization restricts the representational flexibility of the model, especially for cross-variable correlations when the number of categories for categorical variables or the number of levels for ordinal variables are large.
In Sec.~\ref{sec:numerical_validation}, we numerically validate the representational flexibility of the model and investigate the sampling distribution of the maximum likelihood estimator using artificial data.
We also compare the representational flexibility of the proposed model with the log-linear model and the latent class model using a real dataset.
Section~\ref{sec:conclusion} is devoted to conclusions.

In some literature, categorical variables are defined as variables encompassing both nominal and ordinal scales~\citep{Agresti2012, Agresti2010}.
However, in this paper, we distinguish between the two; variables with nominal scales are specifically referred to as ``categorical variables,'' and those with ordinal scales as ``ordinal variables.''
The set of all $P_0$-matrices, that is, a matrix whose principal minors are all nonnegative, is denoted by $\mathcal{P}_0$.

\section{Parametrization for categorical and ordinal variables} \label{sec:result}
In this section, we first summarize the results of the paper, then explain the rationale behind them.
We present a structure of the parameter matrix of the Grassmann distribution that realizes the co-occurrence patterns of the dummy variables for categorical and ordinal variables, and provide concrete examples.
Then, we introduce a diagonally dominant parametrization to ensure nonnegative probabilities that is compatible with the proposed parameter structure.

\subsection{Structure of the parameter matrix}
First, we represent categorical and ordinal variables as combinations of binary variables.
For categorical variables, we use dummy encoding.
We represent a categorical variable with $(q+1)$ categories, $\xi \in \{0,1,2,\dots, q \}$, by a set of $q$ dummy variables, $y_l \in \{0, 1\}, \, (l=1,2,\dots, q)$, where at most one dummy variable is equal to 1, and all others are 0.
In the dummy encoding, we exclude a dummy variable for the base category.
For ordinal variables, we use forward cumulative coding.
We represent an ordinal variable with $(q+1)$ levels, $\xi \in \{0,1,2,\dots, q\}$, by a set of $q$ dummy variables, $y_l \in \{0, 1\}, \, (l=1, 2, \dots, q)$, where each dummy variable indicates whether the ordinal variable is greater than or equal to one, two, etc.
For example, for an ordinal variable with four levels, the value $\xi=2$ is represented as $y_1=1$, $y_2=1$ and $y_3=0$, since $\xi$ is greater than or equal to 1 and 2, but less than 3.
The dummy encoding for a categorical variable with four categories and forward cumulative coding for an ordinal variable with four levels are shown in Table~\ref{table:dummy_encoding}.
Note that for both categorical and ordinal variables, a single dummy variable is not meaningful.
Instead, a set of multiple dummy variables corresponding to the same categorical or ordinal variable is required.
Binary variables can be treated as categorical variables with two categories, or as ordinal variables with two levels.
We, therefore, treat binary variables as a special case of categorical variables.

\begin{table}[hbtp]
    \caption{Dummy encoding of a categorical variable (left) and an ordinal variable (right).}
    \label{table:dummy_encoding}
    \begin{tabular}{cc}
    \begin{minipage}{0.41\hsize}
    \centering
    \begin{tabular}{c|ccc}
        \hline
        $\xi$ & $y_1$ & $y_2$ & $y_3$ \\
        \hline
        $\xi=0$ & 0 & 0 & 0 \\
        $\xi=1$ & 1 & 0 & 0 \\
        $\xi=2$ & 0 & 1 & 0 \\
        $\xi=3$ & 0 & 0 & 1 \\
        \hline
    \end{tabular}
\end{minipage}
\hfill
\begin{minipage}{0.41\hsize}
    \centering
    \begin{tabular}{c|ccc}
        \hline
        $\xi$ & $y_1$ & $y_2$ & $y_3$ \\
        \hline 
        $\xi=0$ & 0 & 0 & 0 \\
        $\xi=1$ & 1 & 0 & 0 \\
        $\xi=2$ & 1 & 1 & 0 \\
        $\xi=3$ & 1 & 1 & 1 \\
        \hline
    \end{tabular}
\end{minipage}
\end{tabular}
\end{table}

Let us consider the case where there are multiple categorical and ordinal variables.
We denote the total number of categorical and ordinal variables by $f$.
First, we define a notation to represent these variables collectively.
We denote a categorical or ordinal variable by $\xi_j, \hspace{0.1cm} (j=1,2,\dots, f)$.
We assign an integer $q_j$ to each variable $\xi_j$ so that $|q_j| + 1$ denotes the number of categories of the categorical variable or the number of levels of the ordinal variable.
If the sign of $q_j$ is positive, the variable $\xi_j$ is treated as a categorical variable, while the sign of $q_j$ is negative, the variable is treated as an ordinal variable.
In other words, the categorical or ordinal variable $\xi_j$ takes values in $\xi_j \in \{0, 1, 2, \dots, | q_j | \}$.
We denote the index set of dummy variables belonging to the $j$th categorical or ordinal variable by $R_j$.
The categorical variable $\xi_j$ is encoded by the column vector of dummy variables $\mathbf{y}_{R_j}$, where $\mathbf{y}_{R_j}$ is a $|q_j|$-dimensional bit vector in which at most only one element can take the value 1 and all the other elements are 0.
Similarly, the ordinal variable $\xi_j$ is encoded by the column vector of dummy variables $\mathbf{y}_{R_j}$, where
\begin{align}
    \mathbf{y}_{R_j} = [1, 1, \dots, 1, 0, 0, \dots, 0]^T \label{eq:ordinal_dummy}
\end{align}
is a $|q_j|$-dimensional bit vector in which all elements that take the value 1 are aligned to the left.
The dummy vectors of categorical and ordinal variables are combined into a single dummy vector $\mathbf{y}$, 
\begin{align}
    \mathbf{y} = [\mathbf{y}_{R_1}; \mathbf{y}_{R_2}; \, \dots \, ; \mathbf{y}_{R_f}],
\end{align}
where the semicolon denotes the vertical concatenation of the column vectors, $[\mathbf{y}_{R_1}; \, \mathbf{y}_{R_2}] \equiv [\mathbf{y}_{R_1}^T, \, \mathbf{y}_{R_2}^T]^T$.
The dimension of the combined dummy vector $\mathbf{y}$ is $q = |q_{1}| + |q_{2}| + \cdots + |q_{f}|$.
We denote the set of all indices $\{1,2,\dots, q\}$ by $R$.
The indexed notation $\mathbf{y}_{R_j}$ denotes the subvector of $\mathbf{y}$ consisting of the components indexed by $R_j$.
We sometimes denote the complement of a set $R_j$ with an overbar, $\bar{R}_j \equiv R \setminus R_j$.

We consider modeling these dummy variables using the Grassmann distribution.
The definition of the Grassmann distribution used in this paper differs from those used in the previous studies~\citep{Arai2021, Arai2022}.
This change of definition is intended to accommodate the parameter matrix of the Grassmann distribution which realizes the co-occurrence probabilities required for categorical and ordinal dummy variables.
The expressions for the Grassmann distribution with modified definition are summarized in Appendix~\ref{sec:inverted_grassmann_distribution}.
Hereafter, we denote the Grassmann distribution used in this paper by $\mathcal{G}^{(1)}$.

Direct application of the Grassmann distribution to these dummy variables is not appropriate.
In the dummy encoding of categorical variables, multiple dummy variables do not take the value 1 simultaneously, and at most one of the dummy variables takes the value 1.
In the forward cumulative coding of ordinal variables, the elements of the dummy vector that take the value 1 are aligned to the left as in Eq.~(\ref{eq:ordinal_dummy}).
In other words, it is not allowed that the lower-level dummy variables take the value 0 while higher-level dummy variables take the value 1.
For example, $y_{l-1}=0$ and $y_l=1$ cannot occur simultaneously.
However, straightforward application of the Grassmann distribution assigns nonzero probabilities to occurrence patterns that are incompatible with the encodings in Table~\ref{table:dummy_encoding}.

To satisfy the co-occurrence probabilities required for categorical and ordinal variables, it is sufficient to follow the next proposition.
\begin{proposition} \label{proposition:categorical_ordinal}
    Let $L_{R_j}$ denote the lower shift matrix of order $|q_j|$ with ones only on the subdiagonal and all the other elements equal to zero as defined by
    \begin{align}
        L_{R_j} = 
        \begin{bmatrix}
            0 & 0 &  0 & \cdots & 0 \\
            1 & 0 &  0 & \cdots & 0 \\
            0 & 1 & \ddots & \ddots & \vdots \\
            \vdots & \ddots & \ddots & 0 & 0 \\
            0 & 0 & 0 &  1 & 0
        \end{bmatrix}. \label{eq:tril_matrix}
    \end{align}
    We consider a $q \times q$ matrix $\Lambda$ such that the row block $(\Lambda - I)_{R_j R}$ has the following form,
    \begin{align}
        & (\Lambda - I)_{R_j R} \notag \\
        &= \begin{cases}
        \begin{bmatrix} (\bm{\lambda}^{(j)})^T \\ \vdots \\ (\bm{\lambda}^{(j)})^T \end{bmatrix}, \hspace{0.2cm}  \{ j \in \{1,2,\dots, f\} \mid  \; q_j > 0 \}, \vspace{0.2cm} \\ 
        \begin{bmatrix} (\bm{\lambda}^{(j)})^T \\ \bm{0}^T \\ \vdots \\ \bm{0}^T \end{bmatrix} 
        - \begin{bmatrix}
            O_{R_j R_1}, & \dots, & O_{R_j R_{j-1}}, & L_{R_j}, & O_{R_j R_{j+1}}, & \dots, & O_{R_j R_f}
        \end{bmatrix},
        \hspace{0.2cm} \{ j \in \{1,2,\dots, f\} \mid  \; q_j < 0 \},
        \end{cases}
        \label{eq:main_result}
    \end{align}
    where $I$ is an identity matrix, the indexed notation $\Lambda_{R_j R}$ expresses the submatrix of $\Lambda$, with its rows and columns consisting of the index sets $R_j$ and $R$, respectively, $\bm{\lambda}^{(j)}$ are $q$-dimensional column vectors, and each subvector $\bm{\lambda}_{R_j}^{(j)}$ has nonnegative elements, $O$ is a zero matrix with all elements equal to zero, in particular, $O_{R_j R}$ is a zero matrix of size $|q_{j}| \times q$.
    If the Grassmann distribution is parametrized so that
    \begin{align}
        p(\mathbf{y}) = \mathcal{G}^{(1)}(\mathbf{y} \mid \Sigma=\Lambda^{-1}),
    \end{align}
    then, the probability that multiple dummy variables belonging to the same categorical variable simultaneously take the value 1, or that a lower-level dummy variable belonging to an ordinal variable takes the value 0 while a higher-level dummy variable corresponding to the same ordinal variable takes the value 1, is zero.
\end{proposition}

\begin{proof}
It is sufficient to focus on the row block of the matrix $\Sigma=\Lambda^{-1}$ indexed by $R_j$ which belongs to a particular categorical or ordinal variable.
To keep the notation concise, we permute the index set $R_j$ to the upper left corner by changing the ordering of the variables, without loss of generality.
Then, in the parametrization of the above proposition, the inverse matrix $\Sigma=\Lambda^{-1}$ is given by
\begin{align}
	\Lambda^{-1} =&
	\begin{bmatrix}
		\Lambda_{R_j | \bar{R}_j}^{-1} & - \Lambda_{R_j | \bar{R}_j}^{-1} \Lambda_{R_j \bar{R}_j} \Lambda_{\bar{R}_j \bar{R}_j}^{-1} \\
		- \Lambda_{\bar{R}_j | R_j}^{-1} \Lambda_{\bar{R}_j R_j} \Lambda_{R_j R_j}^{-1} & \Lambda_{\bar{R}_j | R_j}^{-1}
	\end{bmatrix}
	= \begin{bmatrix}
		\Sigma_{R_j R_j} & \Sigma_{R_j \bar{R}_j} \\
		\Sigma_{\bar{R}_j R_j} & \Sigma_{\bar{R}_j \bar{R}_j}
	\end{bmatrix},
\end{align}
where we have used the partitioned inverse formula, Eq.~(\ref{eq:inverse_of_a_partitioned_matrix}).
We will divide the cases into those in which the index set $R_j$ belongs to a categorical variable and those in which $R_j$ belongs to an ordinal variable.

When the index set $R_j$ corresponds to a categorical variable, the Schur complement $\Lambda_{R_j | \bar{R}_j}$ is given by
\begin{align}
	\Lambda_{R_j | \bar{R}_j} = I + (\Lambda - I)_{R_j R_j} - \Lambda_{R_j \bar{R}_j} \Lambda_{\bar{R}_j \bar{R}_j}^{-1} \Lambda_{\bar{R}_j R_j}.
\end{align}
Since $(\Lambda - I)_{R_j R_j}$ and $\Lambda_{R_j \bar{R}_j}$ are matrices whose row vectors are identical, the Schur complement $\Lambda_{R_j | \bar{R}_j}$ consists of an identity matrix plus a matrix whose rows are all identical.
Then, in general, the matrix $\Lambda_{R_j | \bar{R}_j}$ can be represented by
\begin{align}
	\Lambda_{R_j | \bar{R}_j} = I - W V^T, \hspace{0.3cm}
	W = \begin{bmatrix} \mathbf{w}^T \\ \vdots \\ \mathbf{w}^T \end{bmatrix},
\end{align}
where $\mathbf{w}$ is a column vector of arbitrary dimension, $W$ is a matrix with $q_j$ rows, and $V$ is a matrix that has the same dimension as $W$.
From the Woodbury matrix identity or the matrix inversion lemma, Eq.~(\ref{eq:Woodbury}), the inverse of the Schur complement is given by
\begin{align}
	\Lambda_{R_j | \bar{R}_j}^{-1} =&
	I + W (I - V^T W)^{-1} V^T = \Sigma_{R_j R_j}.
\end{align}
That is, the inverse matrix of an identity matrix plus a matrix whose rows are all identical is also an identity matrix plus a matrix whose rows are all identical.
Furthermore, since the $R_j \bar{R}_j$ block of the matrix $\Sigma$,
\begin{align}
	\Sigma_{R_j \bar{R}_j} =& - \Lambda_{R_j | \bar{R}_j}^{-1} \Lambda_{R_j \bar{R}_j} \Lambda_{\bar{R}_j \bar{R}_j}^{-1}, \notag \\
	=&
	- \Lambda_{R_j \bar{R}_j} \Lambda_{\bar{R}_j \bar{R}_j}^{-1} - W (I - V^T W)^{-1} V^T \Lambda_{R_j \bar{R}_j} \Lambda_{\bar{R}_j \bar{R}_j}^{-1},
\end{align}
is a matrix whose row vectors are identical, the $R_j$ row block of the matrix $\Sigma$ has the following form with a parameter $\gamma_s, \hspace{0.1cm} (s=1,2,\dots,q)$:
\begin{align}
	\Sigma_{R_j R} =
	\begin{bmatrix}
		\gamma_1 + 1 & \gamma_2 & \gamma_3 & \gamma_4 & \cdots & \gamma_q \\
		\gamma_1 & \gamma_2 + 1 & \gamma_3 & \gamma_4 & \cdots & \gamma_q \\
		\gamma_1 & \gamma_2 & \gamma_3 + 1 & \gamma_4 & \cdots & \gamma_q \\
		\vdots        &     \vdots        &     \vdots       & \vdots  &     \cdots     &      \vdots     \\
	\end{bmatrix}.
\end{align}
When two dummy variables in the index set $R_j$ are simultaneously observed to be 1, the corresponding rows of the flipped parameter matrix $\tilde{\Sigma}$, defined in Eq.~(\ref{eq:grassmann_joint_Sigma}), become identical.
Therefore, it follows that the determinant of $\tilde{\Sigma}$ becomes exactly zero, which implies that the undesired occurrence probabilities associated with categorical variables are exactly zero.

When the index set $R_j$ corresponds to an ordinal variable, the Schur complement $\Lambda_{R_j | \bar{R}_j}$ is, in general, written as
\begin{align}
	\Lambda_{R_j | \bar{R}_j} =& I - L - \tilde{W} V^T, \hspace{0.3cm}
	\tilde{W} = \begin{bmatrix} \mathbf{w}^T \\ \bm{0}^T \\ \vdots \\ \bm{0}^T \end{bmatrix},
\end{align}
where $\mathbf{w}$ is again a column vector of arbitrary dimension, $\tilde{W}$ is a matrix with $|q_j|$ rows, and $V$ is a matrix whose dimension is the same as $\tilde{W}$.
Here, we note that by row reduction the inverse of an identity matrix minus lower shift matrix $I - L$ becomes a unit lower triangular matrix, a lower triangular matrix whose nonzero elements are all equal to one, as follows:
\begin{align}
	(I - L)^{-1} =
	\begin{bmatrix}
        1 & 0 & 0 & \cdots & 0 \\
        -1 & 1 & 0 & \ddots & \vdots \\
        0 & -1 & \ddots & \ddots & 0 \\
        \vdots & \ddots & \ddots & 1 & 0 \\
        0 & \cdots & 0 & -1 & 1
	\end{bmatrix}^{-1} 
	=
	\begin{bmatrix}
        1 & 0 & \cdots & 0 \\
        1 & 1 & \ddots & \vdots \\
        \vdots & \vdots & \ddots & 0 \\
        1 & 1 & \cdots & 1
	\end{bmatrix}.
\end{align}
Again, from the Woodbury matrix identity, the inverse matrix of the Schur complement $\Lambda_{R_j | \bar{R}_j}$ is given by
\begin{align}
	\Lambda_{R_j | \bar{R}_j}^{-1} =&
	 (I - L)^{-1} + (I - L)^{-1} \tilde{W} \bigl[ I - V^T (I -L)^{-1} \tilde{W} \bigr]^{-1} V^T (I - L)^{-1}, \notag \\
	=&
	(I - L)^{-1} + W \bigl[ I - V^T W \bigr]^{-1} V^T (I - L)^{-1}, \hspace{0.3cm} W = \begin{bmatrix} \mathbf{w}^T \\ \vdots \\ \mathbf{w}^T \end{bmatrix},
\end{align}
where we have used the relation $(I - L)^{-1} \tilde{W} = W$.
Furthermore, since $(I - L)^{-1} \Lambda_{R_j \bar{R}_j}$ is a matrix whose rows are all identical, the $R_j \bar{R}_j$ block of the matrix $\Sigma$,
\begin{align}
	\Sigma_{R_j \bar{R}_j} =&
	 - \Lambda_{R_j | \bar{R}_j}^{-1} \Lambda_{R_j \bar{R}_j} \Lambda_{\bar{R}_j \bar{R}_j}^{-1}, \notag \\
	=&
	- (I - L)^{-1} \Lambda_{R_j \bar{R}_j} \Lambda_{\bar{R}_j \bar{R}_j}^{-1} - W (I - V^T W)^{-1} V^T (I - L)^{-1} \Lambda_{R_j \bar{R}_j} \Lambda_{\bar{R}_j \bar{R}_j}^{-1},
\end{align}
is also a matrix whose rows are all identical.
Then, the $R_j$ row block of the matrix $\Sigma$ has the following form with a parameter $\gamma_s, \hspace{0.1cm} (s=1,2,\dots,q)$:
\begin{align}
	\Sigma_{R_j R} =
	\begin{bmatrix}
		\gamma_1 + 1 & \gamma_2 & \gamma_3 & \gamma_4 & \cdots & \gamma_q \\
		\gamma_1 + 1 & \gamma_2 + 1 & \gamma_3 & \gamma_4 & \cdots & \gamma_q \\
		\gamma_1 + 1 & \gamma_2 + 1 & \gamma_3 + 1 & \gamma_4 & \cdots & \gamma_q \\
		\vdots       &    \vdots      &     \vdots      &     \vdots  &     \cdots     &     \vdots      \\
	\end{bmatrix}.
\end{align}
Therefore, when the lower-level dummy variable is observed to be 0 while the higher-level dummy variable is observed to be 1, the corresponding rows of the flipped parameter matrix $\tilde{\Sigma}$, defined in Eq.~(\ref{eq:grassmann_joint_Sigma}), become identical.
Consequently, it follows that the determinant of $\tilde{\Sigma}$ becomes exactly zero, which implies that the undesired occurrence probabilities associated with ordinal variables are exactly zero.
This completes the proof.
\end{proof}

\subsection{Examples of the structured parametrization}
\subsubsection{Single categorical variable}
First, we consider an example of occurrence probabilities for a categorical variable.
Let us consider a specific example of four categories, i.e., $q=3$.
In this case, Proposition~\ref{proposition:categorical_ordinal} proposes the following parametrization:
\begin{align}
    \Lambda - I =
    \begin{bmatrix}
        \Lambda_{11}-1 & \Lambda_{22} - 1 & \Lambda_{33} - 1 \\
        \Lambda_{11} - 1 & \Lambda_{22}-1 & \Lambda_{33} - 1 \\
        \Lambda_{11} - 1 & \Lambda_{22} - 1 & \Lambda_{33}-1
    \end{bmatrix}. \label{eq:quasi_diagonal_categorical}
\end{align}
Then, the inverse matrix is given by
\begin{align}
    \Sigma =
    \begin{bmatrix}
        \Sigma_{11} & \Sigma_{22} - 1 & \Sigma_{33} - 1  \\
        \Sigma_{11} - 1 & \Sigma_{22} & \Sigma_{33} - 1  \\
        \Sigma_{11} - 1 & \Sigma_{22} - 1 & \Sigma_{33}
    \end{bmatrix}. 
\end{align}
In the above parametrization, only the four occurrence probabilities listed in Table~\ref{table:dummy_encoding} become nonzero.
The four nonzero probabilities of a categorical variable can be expressed as the following categorical distribution $\mathrm{Cat}(\mathbf{y}| \mathbf{b} )$:
\begin{align}
    p(\xi) = & \mathrm{Cat}(\mathbf{y}| \mathbf{b} ) \equiv \frac{\exp(\mathbf{y}^T \mathbf{b})}{1 + \sum_{l=1}^{q} \exp(b_l)},
    \hspace{0.5cm}
    b_l = \log(\Lambda_{ll} - 1), \hspace{0.3cm} l \in \{1, 2, 3\}.
\end{align}

\subsubsection{Single ordinal variable}
Next, we consider an example of occurrence probabilities for an ordinal variable.
Let us consider a specific example of four levels, i.e., $q=3$.
In this case, Proposition~\ref{proposition:categorical_ordinal} proposes the following parametrization:
\begin{align}
    \Lambda - I=
    \begin{bmatrix}
        \Lambda_{11} -1 & \Lambda_{12} & \Lambda_{13} \\
        -1 & 0 & 0 \\
        0 & -1 & 0
    \end{bmatrix}. \label{eq:quasi_diagonal_ordinal}
\end{align}
Then, the inverse matrix is given by
\begin{align}
    \Sigma =
    \begin{bmatrix}
        \Sigma_{11} & \Sigma_{22} - 1 & \Sigma_{33} - 1 \\
        \Sigma_{11} & \Sigma_{22} & \Sigma_{33} - 1 \\
        \Sigma_{11} & \Sigma_{22} & \Sigma_{33} \\
    \end{bmatrix}.
\end{align}
In the above parametrization, only the four different occurrence probabilities as represented in Table~\ref{table:dummy_encoding} become nonzero.
For all principal minors of the matrix $\Lambda - I$ to be nonnegative, their elements must satisfy $\Lambda_{11}-1 \ge 0$, $\Lambda_{12} \ge 0$ and $\Lambda_{13} \ge 0$.
The probabilities of an ordinal variable taking on nonzero values can be modeled using the ordinal distribution $\mathrm{Ord}(\mathbf{y} | \mathbf{b})$, which is defined as follows:
\begin{align}
    p(\xi) = & \mathrm{Ord}(\mathbf{y} | \mathbf{b}) \equiv \frac{\exp(\mathbf{y}^T \mathbf{b})}{1 + \sum_{l=1}^{q} \prod_{m=1}^l \exp(b_m)},
    \hspace{0.4cm}
    \Lambda_{11}-1 = e^{b_1}, \hspace{0.2cm} \Lambda_{12} = e^{b_1 + b_2}, \hspace{0.2cm}
    \Lambda_{13} = e^{b_1 + b_2 + b_3}.
\end{align}

\subsubsection{Mixed variables of categorical and ordinal type} \label{subsubsec:mixed_type}
Finally, we consider an example of occurrence probabilities for mixed variables consisting of one categorical variable and one ordinal variable, each with four categories and four levels, respectively.
In this case, Proposition~\ref{proposition:categorical_ordinal} proposes the following parametrization:
\begin{align}
    \Lambda - I =&
    {
    \renewcommand{\arraystretch}{1.25}
    \left[
    \begin{array}{ccc|ccc}
        \Lambda_{11} - 1 & \Lambda_{22} - 1 & \Lambda_{33} - 1 &
	\Lambda_{14} & \Lambda_{15} & \Lambda_{16} \\
        \Lambda_{11} - 1 & \Lambda_{22} - 1 & \Lambda_{33} - 1 &
	\Lambda_{14} & \Lambda_{15} & \Lambda_{16} \\
        \Lambda_{11} - 1 & \Lambda_{22} - 1 & \Lambda_{33} - 1 &
	\Lambda_{14} & \Lambda_{15} & \Lambda_{16} \\ \hline
	\Lambda_{41} & \Lambda_{42} & \Lambda_{43} & 
	\Lambda_{44}-1 & \Lambda_{45} & \Lambda_{46} \\
	0 & 0 & 0 & -1 & 0 & 0 \\
	0 & 0 & 0 & 0 & -1 & 0
    \end{array}
    \right]
    },
\end{align}
where $\Lambda_{44}-1 \ge 0$, $\Lambda_{45} \ge 0$ and $\Lambda_{46} \ge 0$.
Then, the inverse matrix is given by
\begin{align}
    \Sigma =&
    {
    \renewcommand{\arraystretch}{1.25}
    \left[
    \begin{array}{ccc|ccc}
        \Sigma_{11} & \Sigma_{22} - 1 & \Sigma_{33} - 1 &
	\Sigma_{14} & \Sigma_{15} & \Sigma_{16} \\
        \Sigma_{11} - 1 & \Sigma_{22} & \Sigma_{33} - 1 &
	\Sigma_{14} & \Sigma_{15} & \Sigma_{16} \\
        \Sigma_{11} - 1 & \Sigma_{22} - 1 & \Sigma_{33} &
	\Sigma_{14} & \Sigma_{15} & \Sigma_{16} \\ \hline
	\Sigma_{41} & \Sigma_{42} & \Sigma_{43} & \Sigma_{44} & \Sigma_{55}-1 & \Sigma_{66}-1 \\
	\Sigma_{41} & \Sigma_{42} & \Sigma_{43} & \Sigma_{44} & \Sigma_{55} & \Sigma_{66}-1 \\
	\Sigma_{41} & \Sigma_{42} & \Sigma_{43} & \Sigma_{44} & \Sigma_{55} & \Sigma_{66} 
    \end{array}
    \right]
    }. \label{eq:example_mixed}
\end{align}
From Eq.~(\ref{eq:example_mixed}), we see the limitation of the representational flexibility of the proposed model.
The proposed model lacks sufficient parameters to freely describe the correlations between dummy variables across different categorical and/or ordinal variables.
In the Grassmann distribution, the correlation between dummy variables is proportional to $\rho_{ij} \propto - \Sigma_{ij} \Sigma_{ji}$. 
In this example, there are nine possible correlations of the dummy variables across different categorical and ordinal variables, but there are only six independent parameters.
Therefore, the model cannot express all nine correlations exactly.
This limitation on the representational flexibility becomes more pronounced as the numbers of categories or levels increases, as numerically verified in Sec.~\ref{sec:numerical_validation}.

\subsection{Extended diagonally dominant parametrization}
\subsubsection{Parameter estimation}
Here, we discuss a method of parameter estimation from observed data $\mathcal{D} \equiv \{\mathbf{y}_1, \mathbf{y}_2, \dots, \mathbf{y}_N \}$.
A common method for parameter estimation is maximum likelihood estimation.
In our model, the log-likelihood function is expressed as
\begin{align}
    l(\Sigma \mid \mathcal{D}) =& \sum_{i=1}^N \log p(\mathbf{y}_i \mid \Sigma), \notag \\
    =& N \sum_{R_1 \subseteq R} \frac{n_{R_1}}{N} \log p(\mathbf{y} = \bm{1}_{R_1} \mid \Sigma), \notag \\
    \equiv & N \sum_{R_1 \subseteq R} \hat{\pi}_{R_1} \log \pi_{R_1}(\Sigma), \label{eq:likelihood_function}
\end{align}
where $\bm{1}_{R_1}$ is a $q$-dimensional bit vector with each element taking a value 0 or 1,
\begin{align}
    [\bm{1}_{R_1}]_s \equiv
    \begin{cases} 1, \hspace{0.5cm} \text{if} \; \; s \in R_1, \\ 
        0, \hspace{0.5cm} \text{if} \; \; s \in R_0,
    \end{cases} \; \; s=1, 2, \dots, q, \label{eq:bit_vector}
\end{align}
and $n_{R_1}$ denotes the count of observations where the observed dummy variable $\mathbf{y}$ is equal to $\bm{1}_{R_1}$.
Of course the count $n_{R_1}$ satisfies $\sum_{R_1 \subseteq R} n_{R_1} = N$.
In other words, the negative log-likelihood is expressed as the cross-entropy between the empirical joint distribution $\hat{\pi}_{R_1}$ and the model's distribution $\pi_{R_1}(\Sigma) \equiv p(\mathbf{y}_{R_1} = \bm{1}, \mathbf{y}_{R_0} = \bm{0} \mid \Sigma)$.
Since it is difficult to solve the log-likelihood maximization analytically, one has to resort to numerical calculations.
For our model to make sense as a probability distribution, all the joint probabilities generated by the model must take nonnegative values.
This is equivalent to requiring $\Lambda - I$ to be a $P_0$-matrix, which means all principal minors of $\Lambda - I$ are nonnegative.
Although the parametrization of the previous subsections realizes the co-occurrence probabilities required for categorical and ordinal variables, the condition that all joint probabilities are nonnegative is not guaranteed.
If we naively parametrize the Grassmann distribution as in Proposition~\ref{proposition:categorical_ordinal}, Eq.~(\ref{eq:main_result}), this can result in negative probabilities.

\subsubsection{Variant of the diagonally dominant parametrization}
A method to ensure that all probabilities of the binary Grassmann distribution are nonnegative is the diagonally dominant parametrization~\citep{Arai2021}.
In this parametrization, $\Lambda - I$ is expressed as the product of a strictly row diagonally dominant matrix $C$ and a diagonally dominant matrix $B$ both with positive diagonal entries as $\Lambda - I =B C^{-1}$.
From Eq.~(\ref{eq:likelihood_function}), we see that the computational complexity of maximum likelihood estimation in the diagonally dominant parametrization is proportional to the number of unique observed states, not to the number of all possible states as is required in the case of the multivariate Bernoulli distribution.
However, the conventional diagonally dominant parametrization cannot accommodate the structure of parameter matrices $\Sigma$ and $\Lambda$ to realize the co-occurrence probabilities required for categorical and ordinal variables.
This is because, for example, for categorical variables, certain rows of $\Lambda - I$ must be identical, but this is incompatible with the diagonal dominancy of $B$ when the number of categories is greater than three. 
Therefore, we propose a variant of the diagonally dominant parametrization described by the following proposition.

\begin{proposition} \label{proposition:BC_parametrization}
    If we parametrize the parameter $\Lambda - I$ of the Grassmann distribution $\mathcal{G}^{(1)}(\Sigma=\Lambda^{-1})$ as $B C^{-1}$ with a $q \times q$ matrix $B$ and $q \times q$ invertible matrix $C$ 
    such that, for $q_j = 1$, corresponding to a binary variable, $B$ is a row diagonally dominant matrix and $C$ is a strictly row diagonally dominant matrix both with positive diagonal entries,
    \begin{align}
        & b_{ss} \ge \sum_{\substack{t=1 \\ t \ne s}}^q |b_{st}|, \hspace{0.5cm} c_{ss} > \sum_{\substack{t=1 \\ t \ne s}}^q |c_{st}|, \hspace{0.3cm} s \in R_j, \label{eq:BC_binary_condition}
    \end{align}
    for $q_j > 1$, corresponding to a categorical variable,
    \begin{align}
        &
        b_{ss} > \sum_{\substack{t=1 \\ t \ne s}}^q |b_{st}|, \hspace{0.3cm} s = \min(R_j),
        \hspace{0.7cm}
        b_{st} = b_{\min(R_j) t}, \hspace{0.3cm} t \in R, \hspace{0.3cm} s \ne \min(R_j), \\
        &
        c_{st} =
        \begin{cases} -c_{ss} < 0, \hspace{0.3cm} t = s + 1, \\
                      0, \hspace{0.3cm} t \notin \{s, s+1 \},
        \end{cases}
        s \ne \max(R_j),  
        \hspace{0.8cm}
        c_{ss} > \sum_{\substack{t=1 \\ t \ne s}}^q |c_{st}|, \hspace{0.3cm} s = \max(R_j), 
        \notag \\
        \label{eq:BC_categorical_condition}
    \end{align}
    and for $q_j \le - 1$, corresponding to an ordinal variable,
    \begin{align}
        &
        b_{ss} > \sum_{\substack{t=1 \\ t \ne s}}^q | b_{st}|, \hspace{0.3cm} s = \min(R_j), \hspace{0.8cm} 
        b_{st} = - c_{(s-1) t}, \hspace{0.3cm} s \ne \min(R_j), \label{eq:BC_ordinal_condition_1} \\
        &
        c_{st} = 
        \begin{cases}
            - c_{ss} < 0, \hspace{0.3cm} t = s + 1, \\
            0, \hspace{0.3cm} t \notin \{s, s+1\},  \\
        \end{cases} s \ne \max(R_j), \hspace{0.8cm}
        c_{ss} > \sum_{\substack{t=1 \\ t \ne s}}^q | c_{st} | , \hspace{0.3cm} s = \max(R_j), \label{eq:BC_ordinal_condition_2}
    \end{align}
    then, the probability that dummy variables belonging to the same categorical variable simultaneously take the value 1, or that a lower-level dummy variable belonging to an ordinal variable takes the value 0 while a higher-level dummy variable corresponding to the same ordinal variable takes the value 1, is zero, and $B C^{-1} \in \mathcal{P}_0$.
\end{proposition}
\begin{proof}
We first show that the diagonally dominant parametrization, Eqs.~(\ref{eq:BC_binary_condition}, \ref{eq:BC_categorical_condition}, \ref{eq:BC_ordinal_condition_1}, \ref{eq:BC_ordinal_condition_2}), satisfies the condition of Eq.~(\ref{eq:main_result}) in Proposition~\ref{proposition:categorical_ordinal}.
Then, we show that this diagonally dominant parametrization yields a $P_0$-matrix.

For $q_j > 0$, it is trivial that Eqs.~(\ref{eq:BC_binary_condition}, \ref{eq:BC_categorical_condition}) satisfy the condition of Proposition~\ref{proposition:categorical_ordinal}.
For an ordinal variable, the rows of the index set $R_j$ of the matrix $\Lambda-I$ are expressed by a negative shift matrix, except for the first row of the index set $R_j$.
In this case, when $\Lambda - I$ is parametrized as $\Lambda - I = B C^{-1}$, the following equation holds,
\begin{align}
    &
    \begin{bmatrix}
        \Lambda_{11}-1 & \Lambda_{12} & \Lambda_{13} & \Lambda_{14} & \dots \\
        -1 & 0 & 0 & 0 & \dots \\
        0 & -1 & 0 & 0 & \dots \\
        0 & 0 & -1 & 0 & \dots \\
	    \vdots & \vdots & \vdots & \vdots & \ddots
    \end{bmatrix}
    = B C^{-1},
    \;\; \Longleftrightarrow \;\; 
    \begin{bmatrix}
        \Lambda_{11}-1 & \Lambda_{12} & \Lambda_{13} & \Lambda_{14} & \dots \\
        -1 & 0 & 0 & 0 & \dots \\
        0 & -1 & 0 & 0 & \dots \\
        0 & 0 & -1 & 0 & \dots \\
	    \vdots & \vdots & \vdots & \vdots & \ddots
    \end{bmatrix} C = B,
\end{align}
where we have permuted the index set $R_j$ to the upper left corner using the invariance of the determinant of the matrix $\Lambda$ under the permutation of a row and its corresponding column, without loss of generality.
Since multiplying a matrix by a lower shift matrix on the left shifts its row vectors down by one row, e.g.,
\begin{align}
    \begin{bmatrix} 0 & 0 & 0 & 0 \\ 1 & 0 & 0 & 0 \\ 0 & 1 & 0 & 0 \\ 0 & 0 & 1 & 0 \end{bmatrix}
    \begin{bmatrix} \mathbf{c}_1^T \\ \mathbf{c}_2^T \\ \mathbf{c}_3^T \\ \mathbf{c}_4^T \end{bmatrix}
    =
    \begin{bmatrix} \bm{0}^T \\ \mathbf{c}_1^T \\ \mathbf{c}_2^T \\ \mathbf{c}_3^T \end{bmatrix},
\end{align}
the above equation is equivalent to having the rows of the matrix $C$ and $B$ coincide as
\begin{align}
	b_{st} = - c_{(s-1)t}, \hspace{0.3cm} s \ne \min(R_j), \hspace{0.3cm} t \in R.
\end{align}
This equation is a part of Eq.~(\ref{eq:BC_ordinal_condition_1}), and thus satisfies Proposition~\ref{proposition:categorical_ordinal}.

$B C^{-1} \in \mathcal{P}_0$ can be proved by contradiction.
By Theorem~\ref{theorem:sandberg} below, if we assume that $B C^{-1} \notin \mathcal{P}_0$, a positive diagonal matrix $D$ exists such that $\det(B C^{-1} + D) = 0$.
Since
\begin{align}
	\det(B C^{-1} + D) = \det C^{-1} \det(B + D C) \equiv \det C^{-1} \det A,
\end{align}
and $C$ is invertible, $\det(B C^{-1} + D) = 0$ is equivalent to $\det A=0$.
The condition $\det A=0$ is equivalent to the fact that there exists nonzero $q$-dimensional column vector $\mathbf{z}$ such that $A \mathbf{z} = \bm{0}$.
We denote the index of maximum absolute value of the elements $\mathbf{z} = (z_1, z_2, \dots, z_q)^T$ as $s$.
We also denote $E \equiv DC$.

If we assume that $s$ corresponds to the index for binary dummy variables, this assumption contradicts to the strict diagonal dominance of the matrix $A$,
\begin{align}
    \sum_{t=1}^q a_{st} z_t = 0, \hspace{0.3cm} \Longrightarrow \hspace{0.3cm}
    a_{ss} z_s = - \sum_{t \ne s} a_{st} z_t.
\end{align}
Taking the absolute value of both sides yields the following equation,
\begin{align}
    |a_{ss} z_s| =& |a_{ss}| |z_s| = \biggl| - \sum_{t \ne s} a_{st} z_t \biggr| 
    \le \sum_{t \ne s} |a_{st}| |z_t| \le \sum_{t \ne s} |a_{st}| |z_s|, \notag \\
    \Longrightarrow |a_{ss}| \le& \sum_{t \ne s} |a_{st}|. \label{eq:A_condition}
\end{align}
However, it can be shown that the sum of a diagonally dominant matrix and strictly diagonally dominant matrix both with positive diagonal entries is a strictly diagonally dominant matrix.
In fact, $s$th row of the matrix $A = B + D C$ is a strictly diagonally dominant:
\begin{align}
	\sum_{\substack{t=1 \\ t \ne s}}^q |a_{st}| =& \sum_{t \ne s} |b_{st} + d_{ss} c_{st}| \le \sum_{t \ne s} |b_{st}| + d_{ss} \sum_{t \ne s}|c_{st}|, \notag \\
	\le & b_{ss} + d_{ss} \sum_{t \ne s} |c_{st}| < b_{ss} + d_{ss} c_{ss} = a_{ss}.
\end{align}
Therefore, Eq.~(\ref{eq:A_condition}) contradicts to the strict diagonal dominance of the $s$th row of the matrix $A$.

If we assume that $s$ corresponds to the index for ordinal dummy variables, this assumption again leads to a contradiction.
In fact, if we assume that $s$ corresponds to the index $\min(R_j)$ or $\max(R_j)$, this assumption contradicts to the strict diagonal dominance of the matrix $A$ as in the case of the binary variables.
If we assume that $s$ corresponds to the index $\min(R_j) < s < \max(R_j)$, the $s$th element of the equation $A \mathbf{z} = \bm{0}$ leads the following equation,
\begin{align}
    & e_{ss} (z_s - z_{s+1}) + e_{(s-1)(s-1)} (z_s - z_{s-1}) = 0. \label{eq:proof_contradiction_1}
\end{align}
If $z_s - z_{s-1} = 0$, this leads $z_s = z_{s-1} = \dots = z_{\min(R_j)}$, which contradicts to the strict diagonal dominance of the $\min(R_j)$th row of the matrix $B$.
If $z_s - z_{s-1} \ne 0$, the above equation leads the relation:
\begin{align}
    & 0 < \frac{e_{(s-1)(s-1)}}{e_{ss}} \equiv \frac{d_{(s-1)(s-1)} c_{(s-1)(s-1)}}{d_{ss} c_{ss}} = - \frac{(z_s - z_{s+1})}{(z_s - z_{s-1})}. \label{eq:proof_contradiction_2}
\end{align}
The sign of the right hand side of the above equation always be negative, $-\mathrm{sign}(z_s) / \mathrm{sign}(z_s) = -1$, this contradicts the sign of the left hand side.

If we assume that $s$ corresponds to the categorical dummy variables, this assumption again leads to a contradiction.
In fact, if $s$ corresponds to $\min(R_j)$, this contradicts to the strict diagonal dominance of the matrix $A$.
If $s$ corresponds to $\min(R_j) < s < \max(R_j)$, the subtraction of the $(s-1)$th element of the $A \mathbf{z} = \bm{0}$ from the $s$th element leads the same equation as Eq.~(\ref{eq:proof_contradiction_1}):
\begin{align}
    e_{ss} (z_s - z_{s+1}) - e_{(s-1)(s-1)}(z_{s-1} - z_s) = 0.
\end{align}
If $z_s - z_{s-1} = 0$, repeated use of the above expression leads $z_s = z_{s-1} = \dots = z_{\min(R_j)}$, which contradicts to the strict diagonal dominance of the $\min(R_j)$th row of the matrix $B$.
If $z_s - z_{s-1} \ne 0$, the above equation leads the same equation as Eq.~(\ref{eq:proof_contradiction_2}), which contradicts the assumption of the sign of $e_{ss}$.
If $s$ corresponds to $\max(R_j)$, the subtraction of the $(s-1)$th element of the $A \mathbf{z} = \bm{0}$ from the $s$th element leads the equation:
\begin{align}
    & \sum_{t=1}^q e_{st} z_t - e_{(s-1)(s-1)}(z_{s-1} - z_s) = 0.
\end{align}
If $z_s - z_{s-1} = 0$, the above equation contradicts to the strict diagonal dominance of the $\max(R_j)$th row of the matrix $C$.
If $z_s - z_{s-1} \ne 0$, the above equation leads the expression:
\begin{align}
    0 < e_{(s-1)(s-1)} \equiv & d_{(s-1)(s-1)} c_{(s-1)(s-1)}, \notag \\
    =&
    - \frac{d_{ss}}{(z_s - z_{s-1})} \sum_{t=1}^q c_{st} z_t, \notag \\
    =&
    - \frac{d_{ss}}{(z_s - z_{s-1})} \biggl\{ c_{ss} z_s + \sum_{t \ne s} c_{st} z_t \biggr\}, \notag \\
    =&
    - \frac{d_{ss}}{(z_s - z_{s-1})} \biggl\{ \Delta c_{ss} z_s + \sum_{t \ne s} \Bigl( |c_{st}| z_s + c_{st} z_t \Bigr) \biggr\}, \notag \\
    =&
    - \frac{d_{ss}}{(z_s - z_{s-1})} \biggl\{ \Delta c_{ss} z_s + \sum_{t \ne s} |c_{st}| \bigl( z_s + \mathrm{sign}(c_{st}) z_t \bigr) \biggr\},
\end{align}
where $\Delta c_{ss} \equiv c_{ss} - \sum_{t \ne s} |c_{st}| > 0$.
The sign of the right hand side of the above expression is negative, $- \mathrm{sign}(z_s)/ \mathrm{sign}(z_s) = -1$, this contradicts the sign of the left hand side.
Therefore, the assumption $\det A = 0$ leads the contradiction.
Thus, $\det (B C^{-1} + D) \ne 0$ for an arbitrary positive diagonal matrix $D$, and thus, $B C^{-1} \in \mathcal{P}_0$.

Finally, we show that the matrix $C$ is invertible.
When all variables are binary, $C$ is a strictly diagonally dominant matrix and hence invertible.
Therefore, we consider the case in which there is at least one categorical or ordinal variable.
Since simultaneous permutations of rows and columns leave the determinant unchanged, we can, without loss of generality, express $C$ in the following form, with the diagonal block $C_{R_j R_j}$ corresponding to a categorical or ordinal variable $j$ in the upper-left corner:
\begin{align}
    C =&
    {
    \renewcommand{\arraystretch}{1.25}
    \left[
    \begin{array}{c|c}
        C_{R_j R_j} & C_{R_j \bar{R}_j} \\ \hline
        C_{\bar{R}_j R_j} & C_{\bar{R}_j \bar{R}_j}
    \end{array}
    \right]
    }
    =
    {
    \renewcommand{\arraystretch}{1.25}
    \left[
    \begin{array}{cc|c}
        C_{\mathcal{R}_j \mathcal{R}_j} & C_{\mathcal{R}_j \max(R_j)} & C_{\mathcal{R}_j \bar{R}_j} \\
        C_{\max(R_j) \mathcal{R}_j} & C_{\max(R_j) \max(R_j)} & C_{\max(R_j) \bar{R}_j} \\ \hline
        C_{\bar{R}_j \mathcal{R}_j} & C_{\bar{R}_j \max(R_j)} & C_{\bar{R}_j \bar{R}_j}
    \end{array}
    \right]
    }, \notag \\
    =&
    {
    \renewcommand{\arraystretch}{1.25}
    \left[
    \begin{array}{c|cc}
        C_{\mathcal{R}_j \mathcal{R}_j} & C_{\mathcal{R}_j \max(R_j)} & C_{\mathcal{R}_j \bar{R}_j} \\ \hline
        C_{\max(R_j) \mathcal{R}_j} & C_{\max(R_j) \max(R_j)} & C_{\max(R_j) \bar{R}_j} \\
        C_{\bar{R}_j \mathcal{R}_j} & C_{\bar{R}_j \max(R_j)} & C_{\bar{R}_j \bar{R}_j}
    \end{array}
    \right]
    }
    =
    {
    \renewcommand{\arraystretch}{1.25}
    \left[
    \begin{array}{c|c}
        C_{\mathcal{R}_j \mathcal{R}_j} & C_{\mathcal{R}_j \bar{\mathcal{R}}_j} \\ \hline
        C_{\bar{\mathcal{R}}_j \mathcal{R}_j} & C_{\bar{\mathcal{R}}_j \bar{\mathcal{R}}_j}
    \end{array}
    \right]
    }
    =
    \begin{bmatrix}
        C_{\mathcal{R}_j R} \\ C_{\bar{\mathcal{R}}_j R}
    \end{bmatrix},
\end{align}
where we have partitioned the index set $R_j$ corresponding to a categorical or ordinal variable into the last index $\max(R_j)$ and the index set $\mathcal{R}_j$ other than the last index as $R_j = (\mathcal{R}_j, \max(R_j))$.
For illustrative purposes, we consider the case of four categories or four ordinal levels.
However, the following argument applies to any number of categories or ordinal levels.
In this case,
\begin{align}
    C_{\mathcal{R}_j R} =
    \begin{bmatrix} C_{\mathcal{R}_j \mathcal{R}_j} & C_{\mathcal{R}_j \bar{\mathcal{R}}_j} \end{bmatrix}
    =
    \left[
    \begin{array}{ccc|ccc}
        c_{11} & -c_{11} & 0 & 0 & \multicolumn{2}{c}{\multirow{3}{*}{$O_{\mathcal{R}_j \bar{R}_j}$}} \\
        0 & c_{22} & -c_{22} & 0 & \multicolumn{2}{c}{} \\
        0 & 0 & c_{33} & -c_{33} & \multicolumn{2}{c}{}
    \end{array}
    \right],
\end{align}
where $O_{\mathcal{R}_j \bar{R}_j}$ is a zero matrix with the size corresponding to the specified indices.
Since $C_{\mathcal{R}_j \mathcal{R}_j}$ is an upper bidiagonal matrix with nonzero diagonal elements, it is invertible, and by using elementary column operations, its inverse can be expressed as the following upper triangular matrix:
\begin{align}
    C_{\mathcal{R}_j \mathcal{R}_j}^{-1}
    =&
    \begin{bmatrix}
        c_{11} & -c_{11} & 0 \\
        0 & c_{22} & -c_{22} \\
        0 & 0 & c_{33}
    \end{bmatrix}^{-1}
    =
    \begin{bmatrix}
        1/c_{11} & 1/c_{22} & 1/c_{33} \\
        0 & 1/c_{22} & 1/c_{33} \\
        0 & 0 & 1/c_{33}
    \end{bmatrix}.
\end{align}
Then, since the diagonal block $C_{\mathcal{R}_j \mathcal{R}_j}$ is invertible, the determinant of the matrix $C$ can be calculated by the Schur complement formula for the determinant as follows:
\begin{align}
    \det C =&
    \det \left( C_{\mathcal{R}_j \mathcal{R}_j} \right)
    \det \left( C_{\bar{\mathcal{R}}_j \mid \mathcal{R}_j} \right), \notag \\
    =&
    \det \left( C_{\mathcal{R}_j \mathcal{R}_j} \right)
    \det
    \left(
    C_{\bar{\mathcal{R}}_j \bar{\mathcal{R}}_j} - C_{\bar{\mathcal{R}}_j \mathcal{R}_j} 
    C_{\mathcal{R}_j \mathcal{R}_j}^{-1}
    C_{\mathcal{R}_j \bar{\mathcal{R}}_j}
    \right), \notag \\
    =&
    \det \left( C_{\mathcal{R}_j \mathcal{R}_j} \right)
    \det
    \left(
    C_{\bar{\mathcal{R}}_j \bar{\mathcal{R}}_j} - C_{\bar{\mathcal{R}}_j \mathcal{R}_j} 
    \begin{bmatrix}
        1/c_{11} & 1/c_{22} & 1/c_{33} \\
        0 & 1/c_{22} & 1/c_{33} \\
        0 & 0 & 1/c_{33}
    \end{bmatrix}
    \left[
    \begin{array}{ccc}
        0 & \multicolumn{2}{c}{\multirow{3}{*}{$O_{\mathcal{R}_j \bar{R}_j}$}} \\
        0 & \multicolumn{2}{c}{} \\
        -c_{33} & \multicolumn{2}{c}{}
    \end{array}
    \right]
    \right), \notag \\
    =&
    \det \left( C_{\mathcal{R}_j \mathcal{R}_j} \right)
    \det
    \left(
    C_{\bar{\mathcal{R}}_j \bar{\mathcal{R}}_j} + C_{\bar{\mathcal{R}}_j \mathcal{R}_j}
    \left[
    \begin{array}{ccc}
        1 & \multicolumn{2}{c}{\multirow{3}{*}{$O_{\mathcal{R}_j \bar{R}_j}$}} \\
        1 & \multicolumn{2}{c}{} \\
        1 & \multicolumn{2}{c}{}
    \end{array}
    \right]
    \right).
\end{align}
Computing the above Schur complement of $C$ with respect to $C_{\mathcal{R}_j \mathcal{R}_j}$ amounts to adding the sum of the columns of off-diagonal block  $C_{\bar{\mathcal{R}}_j \mathcal{R}_j}$ to the first column of the diagonal block matrix $C_{\bar{\mathcal{R}}_j \bar{\mathcal{R}}_j}$.
When the rows in the indices $\bar{\mathcal{R}}_j$ correspond to the rows with diagonal dominance, the above addition of the off-diagonal elements of $C_{\bar{\mathcal{R}}_j \mathcal{R}_j}$ preserves diagonal dominance.
When the rows in the indices $\bar{\mathcal{R}}_j$ correspond to categorical and ordinal variables other than the last elements within those variables, the corresponding off-diagonal elements in $C_{\bar{\mathcal{R}}_j \mathcal{R}_j}$ are all zero.
Hence, these rows remain unchanged in the Schur complement $C_{\bar{\mathcal{R}}_j \mid \mathcal{R}_j}$.
Therefore, the resulting Schur complement of $C$ with respect to $C_{\mathcal{R}_j \mathcal{R}_j}$ retains the matrix structure described in Proposition 2.
By repeating this process for all categorical and ordinal variables, we eventually obtain the determinant of the final Schur complement consisting entirely of strictly diagonally dominant rows.
The final Schur complement is a strictly diagonally dominant matrix and hence invertible.
Since all the eliminated diagonal blocks are invertible, $C$ is also invertible.

This completes the proof.
\end{proof}

We quote a theorem proven by Sandberg-Willson, Jr.~\citep{Sandberg1969}.
\begin{theorem} \label{theorem:sandberg}
    $A \in \mathcal{P}_0$ is equivalent to $\det(A + D) \ne 0$ for an arbitrary positive diagonal matrix $D$.
\end{theorem}

Another method to ensure nonnegative probabilities is to explicitly impose a nonnegativity constraint on all $2^q$ joint probabilities or introduce a prior distribution on empirical joint probabilities.
In the prior distribution approach, one adopts the maximum a posteriori estimation instead of the maximum likelihood estimation.
To be more specific, one assigns a Dirichlet prior to the vector of model joint probabilities $\pi_{R_1}$, except for the co-occurrence probabilities which are not allowed for categorical and ordinal variables.
Then, the posterior distribution of the model parameters is as follows,
\begin{align}
    p(\Sigma \mid \mathcal{D}) \propto & \; p(\mathcal{D} \mid \Sigma) \; p(\pi(\Sigma) \mid \bm{\alpha}), \notag \\
    =& 
    \frac{1}{B(\bm{\alpha})} \prod_{R_1 \subseteq R} p(\mathbf{y}=\bm{1}_{R_1} \mid \Sigma)^{n_{R_1}} \;\pi_{R_1}(\Sigma)^{\alpha_{R_1} - 1}, \notag \\
    \propto & 
    \prod_{R_1 \subseteq R} p(\mathbf{y}=\bm{1}_{R_1} \mid \Sigma)^{n_{R_1} + \alpha_{R_1} - 1}
\end{align}
where $B(\bm{\alpha})$ is the multivariate beta function and $\alpha_{R_1}$ is the concentration parameter of the Dirichlet distribution.
The use of this prior simply amounts to adding the prior hyper-parameters to the empirical counts, $n_{R_1} \rightarrow n_{R_1} + \alpha_{R_1} - 1$, which is known as pseudo-counts.
Although introducing a constraint condition or a prior distribution to account for the nonnegativity of all possible probabilities serves as a regularization, it also introduces a problem of computational complexity similar to those encountered with the multivariate Bernoulli distribution.
Both approaches are impractical when the number of variables is large, as the number of states required for the log-likelihood calculation increases exponentially.
Therefore, these approaches should be regarded as a temporary solution.

Sections~\ref{subsec:illustrative_example}--\ref{subsec:sampling_behavior} of this paper adopted the direct parametrization of the structure given in Proposition~\ref{proposition:categorical_ordinal} subject to nonnegativity constraints on all joint probabilities, rather than the extended diagonally dominant parametrization given in Proposition~\ref{proposition:BC_parametrization}.
However, as we will see, the extended diagonally dominant parametrization does not compromise the representational flexibility of the model.
The reason for this is that, depending on how it is implemented, the direct parametrization can sometimes be faster, especially when the number of possible states is small.
Since the optimization of the likelihood function is a nonlinear optimization problem with numerous local maxima, it is difficult to obtain the maximum likelihood estimator.
Therefore, in this paper, unless otherwise stated, we generated more than 100 random initial parameter values and numerically solved the optimization problem for each initial parameter, and then adopted the parameter with the highest likelihood value as the optimal solution.

\section{Numerical validation} \label{sec:numerical_validation}
In this section, we numerically validate the proposed model using a real dataset and artificially generated datasets.
Then, we compare the representational flexibility of the proposed model with that of existing models.

\subsection{Illustrative example of the proposed distribution} \label{subsec:illustrative_example}
We first examine whether the proposed model satisfies the zero-probability constraints implied by the dummy encodings and reproduces empirical correlations.
The dataset used in this analysis is the reader data from the \texttt{catdata} package in R (programming language)~\citep{catdata}.
The reader data contain information on the reading behavior of women referring to a specific women's journal and consists of four variables: is the woman a regular reader? ($\text{yes}=1, \, \text{no}=0$), is the woman working? ($\text{yes}=1, \, \text{no}=0$), her age group ($18 \text{--} 29: \text{years} = 1, \,  30 \text{--} 39: \text{years}= 2, \, 40 \text{--} 49: \text{years} = 3$), and her education level from L1 to L4.
The sample size of the dataset is $N=941$.
We considered age group as a categorical variable and education level as an ordinal variable.
We selected one binary variable, working or not (\texttt{working}), one categorical variable, age category (\texttt{age}), and one ordinal variable, education level (\texttt{education}), for analysis.

\begin{figure}[htbp]
    \centering
    \includegraphics[scale=1, pagebox=cropbox, viewport= 0 0 316.563 183.231, clip]{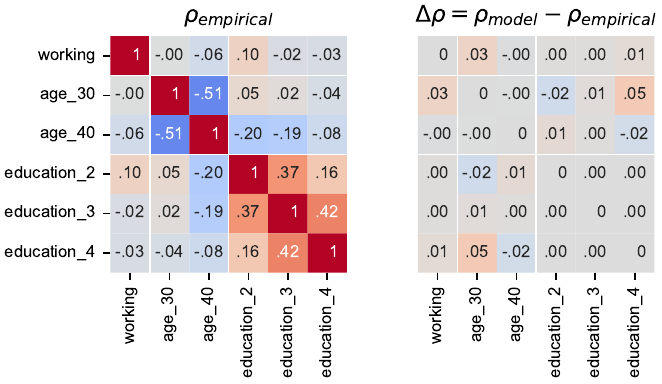}
    \caption{
    Pearson correlation matrix of the reader data, empirically computed from the data (left) and the difference between empirical correlation and the correlation reproduced by the model (right).
    }
    \label{fig:correlation_matrix_reader}
\end{figure}

Model parameters were estimated by maximum likelihood estimation.
The estimated parameters are given by
\begin{align}
    \Lambda - I =& 
    {
    \renewcommand{\arraystretch}{1.25}
    \left[
    \begin{array}{c|cc|ccc}
        0.62 & 0.18 & -0.21 & 1.7 & 0.50 & 0.11 \\ \hline
        -1.2 & 1.7 & 2.6 & 1.7 & 2.3 & 0.95 \\
        -1.2 & 1.7 & 2.6 & 1.7 & 2.3 & 0.95 \\ \hline
        -1.1 & 0.71 & 1.9 & 2.4 & 2.1 & 0.78 \\
        0 & 0 & 0 & -1 & 0 & 0 \\
        0 & 0 & 0 & 0 & -1 & 0 
    \end{array}
    \right]
    }, \\
    \Sigma = &
    {
    \renewcommand{\arraystretch}{1.25}
    \left[
    \begin{array}{c|cc|ccc}
        0.52 & -0.15 & 0.31 & -0.32 & 0.065 & 0.039 \\ \hline
        0.049 & 0.62 & -0.30 & -0.28 & -0.30 & -0.098 \\
        0.049 & -0.38 & 0.70 & -0.28 & -0.30 & -0.098 \\ \hline
        0.069 & 0.019 & -0.13 & 0.22 & -0.32 & -0.08 \\
        0.069 & 0.019 & -0.13 & 0.22 & 0.68 & -0.08 \\
        0.069 & 0.019 & -0.13 & 0.22 & 0.68 & 0.92
    \end{array}
    \right]
    },
\end{align}
where each row and column of the parameter matrix corresponds to the dummy variables \texttt{working}, \texttt{age\_30}, \texttt{age\_40}, \texttt{education\_2}, \texttt{education\_3} and \texttt{education\_4}, in ascending order of indices, and the values are rounded for presentation. 
Although the Grassmann distribution defines $2^6=64$ occurrence probabilities for the dummy variables, all undesired co-occurrence probabilities become exactly zero, while the desirable 12 occurrence probabilities remain nonzero.
The proposed distribution exactly reproduced the empirical means of the dummy variables.
Figure~\ref{fig:correlation_matrix_reader} shows the empirical correlation as well as the correlation reproduced by the model.
We see that the proposed distribution captures the main correlation structure among the dummy variables.

\subsection{Representational flexibility}
Next, we examine the representational flexibility of our model.
We artificially generated datasets for categorical and ordinal variables and checked whether the model can describe the mean and correlation of the generated data.
To generate artificial data for discrete variables, we first generated correlated continuous variables from a multivariate standard normal distribution.
Discrete variables were then obtained by thresholding the continuous variables.
The parameters used for data generation were randomly generated as follows.
We generated the covariance matrix of the multivariate standard normal distribution using spectral decomposition.
First, the eigenvalues of the positive definite symmetric matrix were generated using a Gamma distribution with a shape parameter $\alpha=1$ and a rate parameter $\beta=1$, $\mathrm{Gamma}(\alpha=1, \beta=1)$.
Next, we generated vectors with elements following the standard normal distribution and orthonormalized them using Gram-Schmidt orthonormalization to obtain the orthonormal eigenvectors for each eigenvalue.
Using these eigenvalues and eigenvectors, we formed a positive definite matrix through spectral decomposition.
This positive definite symmetric matrix was then standardized to obtain the covariance matrix of the multivariate standard normal distribution.
The continuous variables generated with this covariance matrix were dichotomized by applying thresholds based on specific percentiles of the standard normal distribution.
The thresholds used for dichotomization to obtain categorical dummy variables were generated from the continuous uniform distribution ranging from zero to one, $\mathrm{Unif}(0,1)$.
We obtained candidates for categorical dummy variables by thresholding the $q_j$ continuous variables based on the generated values.
To convert the resulting dummy variables to categorical ones, we grouped the resulting dummy variables and excluded data points involving the co-occurrence patterns not allowed for the categorical variables.
The ordinal dummy variables were generated by dividing the continuous variables into different levels using specific thresholds.
For ordinal variables, we generated random numbers from the Dirichlet distribution with $|q_j| + 1$ components with each parameter set to one, $\mathrm{Dir}(\bm{\alpha}=(1,\dots,1))$, which corresponds to a uniform distribution over the simplex.
Then, the percentile thresholds of the standard normal distribution were obtained from the cumulative sum of the generated values.
We converted the single continuous variable into an ordinal variable with $|q_j| + 1$ levels using the obtained percentile thresholds, and then represented it by ordinal dummy variables through cumulative dummy encoding.

In this experiment, we generated 1000 sets of artificial data, each with a sample size of 1000.
In each dataset, there are two categorical variables with three and four categories, and two ordinal variables with three and four levels, making a 10-dimensional dataset.
We excluded the artificial data in which any of the dummy variables had an average of either zero or one, since correlation cannot be defined for such data.
The parameters of the model were estimated by maximum likelihood estimation.

\begin{figure}[htbp]
    \centering
    \includegraphics[scale=1, pagebox=cropbox, viewport= 0 0 361.791 192.529, clip]{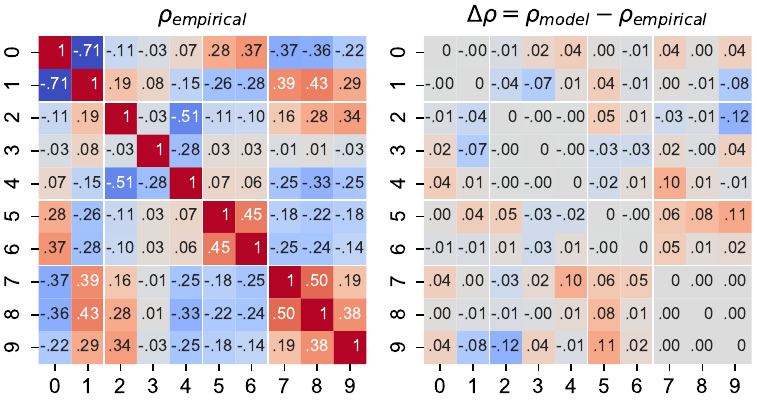}
    \caption{
        Pearson correlation matrix of the artificial data, empirically computed from the data (Left) and the difference between empirical correlation and the correlation reproduced by the model (Right).
    }
    \label{fig:correlation_matrix_artificial}
\end{figure}

\begin{figure}[htbp]
    \centering
    \includegraphics[scale=1, pagebox=cropbox, viewport= 0 0 418.833 304.009, clip]{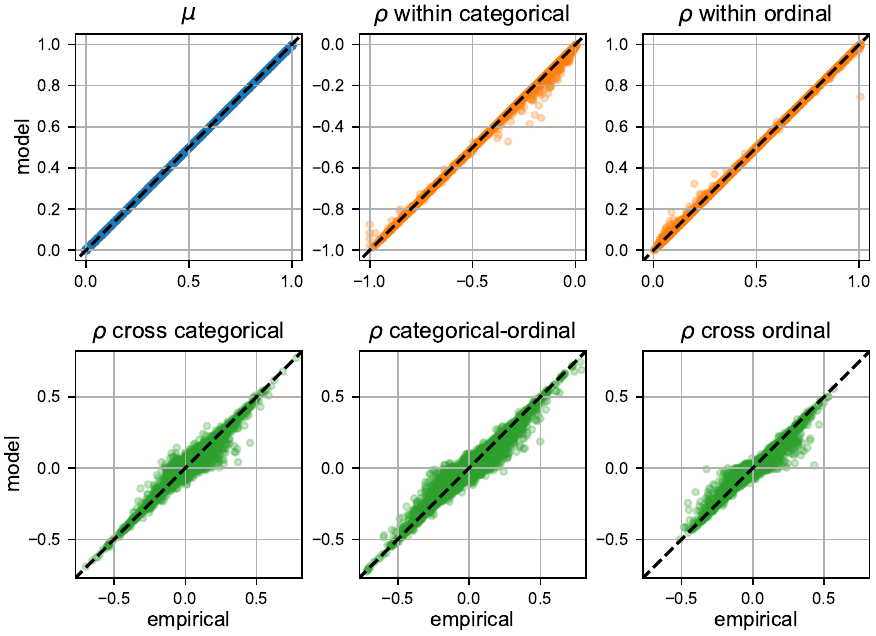}
    \caption{
        Scatter plots of the means $\mu$ and correlations $\rho$ for the empirical distribution against those reproduced by the model.
        The figure titled ``within'' illustrates the correlations between dummy variables belonging to the same categorical or ordinal variable.
        The figure titled ``cross'' illustrates the correlations between dummy variables across different categorical or ordinal variables.
    }
    \label{fig:model_representability}
\end{figure}

Figure~\ref{fig:correlation_matrix_artificial} shows the empirical correlation matrix and the difference between the empirical correlations and those reproduced by the model for one of the artificially generated datasets.
Figure~\ref{fig:model_representability} presents a scatter plot comparing the empirical means and correlations of the generated data with those reproduced by the model.
These plots indicate that our model reproduces the means of the dummy variables exactly and captures the correlation between dummy variables belonging to the same categorical or ordinal variables quite well.
Furthermore, our model also reproduces the correlation between dummy variables across different categorical and ordinal variables reasonably well, although it underestimates the empirical correlations.
This underestimation is consistent with the parameter-count argument in Sec.~\ref{subsubsec:mixed_type}, which suggests limited flexibility for cross-variable associations.

\subsection{Sampling distribution of the estimator} \label{subsec:sampling_behavior}
\begin{figure}[htbp]
    \centering
    \includegraphics[scale=1, pagebox=cropbox, viewport= 0 0 421.254 276.663, clip]{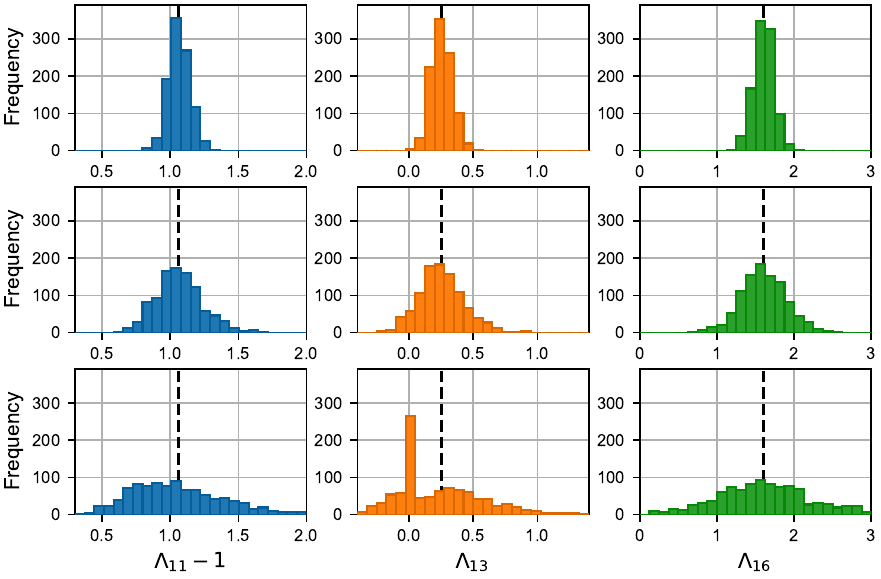}
    \caption{
        Sampling distribution of an estimator for the parameter $\Lambda - I$.
        The true parameters are indicated by the dashed lines.
        The sample sizes are 16000 for the upper row figures, 4000 for the middle row figures, and 1000 for the lower row figures.
    }
    \label{fig:model_sampling_dist}
\end{figure}

In this subsection, we investigate the sampling distribution of the maximum likelihood estimates and examine the consistency of the model parameters.
We generated 1000 artificial datasets with sample sizes of 1000, 4000 and 16000, respectively.
To generate artificial data, we used, as ground truth, one parameter setting previously estimated from an artificial dataset that exhibited strong correlations.
Again, we excluded the artificial data in which any dummy variable had an average of either zero or one, since correlation cannot be defined for such data.
Then, we fitted the model to these datasets by maximum likelihood estimation.
Note that the matrix $\Lambda - I$ has redundant degrees of freedom in constant multiplications and the matrix transposition as shown in Eqs.~(\ref{eq:multiplication_dof}, \ref{eq:transposition_dof}).
To compare the estimated model parameters, we fixed the degrees of freedom.
The degree of freedom related to the matrix transposition is automatically fixed to realize the co-occurrence patterns of categorical and ordinal variables.
To fix the degrees of freedom related to the multiplications by constants, and make the model identifiable, we set the value of the first nonzero element along columns to -1 in each row of the matrix $\Lambda - I$, except the rows corresponding to the dummy variables of the first feature variable, e.g.,
\begin{align}
    \Lambda -I =
    {
    \renewcommand{\arraystretch}{1.25}
    \left[
    \begin{array}{cc|ccc|cc|ccc}
        1.1 & 0.45 & 0.25 & -0.040 & -2.25 & 1.6 & 9.0 & -0.17 & -0.75 & -0.32 \\
        1.1 & 0.45 & 0.25 & -0.040 & -2.25 & 1.6 & 9.0 & -0.17 & -0.75 & -0.32 \\ \hline
        -1 & 0.19 & 0.11 & 0.25 & 12 & 0.70 & -7.2 & -0.074 & 2.9 & 1.5 \\
        -1 & 0.19 & 0.11 & 0.25 & 12 & 0.70 & -7.2 & -0.074 & 2.9 & 1.5 \\
        -1 & 0.19 & 0.11 & 0.25 & 12 & 0.70 & -7.2 & -0.074 & 2.9 & 1.5 \\ \hline
        -1 & -0.043 & 0.035 & -0.0042 & -0.16 & 1.5 & 1.5 & -0.011 & -0.056 & -0.024 \\
        0 & 0 & 0 & 0 & 0 & -1 & 0 & 0 & 0 & 0 \\ \hline
        -1 & 7.7 & -0.22 & 0.062 & 3.3 & 0.19 & -9.0 & 0.75 & 1.1 & 0.45 \\
        0 & 0 & 0 & 0 & 0 & 0 & 0 & -1 & 0 & 0 \\
        0 & 0 & 0 & 0 & 0 & 0 & 0 & 0 & -1 & 0 \\
    \end{array}
    \right]
    }.
\end{align}
Although the Grassmann distribution defines $2^{10} =1024$ occurrence probabilities for the dummy variables in the above parameters, all undesired co-occurrence probabilities become exactly zero, while the desirable $3^2 \times 4^2 =144$ occurrence probabilities remain nonzero.
Figure~\ref{fig:model_sampling_dist} shows the sampling distribution of the estimator for a particular element of $\Lambda$.
We see that as the sample size increases, the estimators converge to the true values, which provides numerical evidence that the estimator appears consistent.
When the sample size is quadrupled, the standard deviation of the estimator's variability is halved.
These results suggest that the parametrization is identifiable after the redundant degrees of freedom are fixed.

\subsection{Comparison with the existing models}
In this subsection, we compare the representational flexibility of the proposed model with that of the existing models.
We adopt a log-linear model with interactions up to second order and the latent class model as the existing models.
Using a real dataset, we examined the goodness-of-fit and the computation time for parameter estimation for each model.
Numerical computations in this subsection were conducted on a personal computer equipped with an Intel Core i7-9700 CPU and 16 GB of RAM, running Windows 11.

We analyzed Big Five personality Test responses included in the Firstborn Personality Scale Responses dataset~\citep{Rohrer2015}.
This data contains questionnaire responses designed to measure five personality traits, with each item rated on a five-point Likert scale.
Specifically, the five traits are extraversion ($\texttt{EXT}$), emotional stability ($\texttt{EST}$), agreeableness ($\texttt{AGR}$), conscientiousness ($\texttt{CSN}$), and openness to experience ($\texttt{OPN}$), with ten questionnaire items for each trait.
We excluded data points with missing values in any responses and data points in which the response of age is greater than hundred.
The sample size after data cleansing was $N = 34,736$.
We split this data into equal sets of training and test data.

We used $\texttt{loglm}$ function from the MASS package~\citep{MASS} for the log-linear model and $\texttt{poLCA}$ function from the poLCA package~\citep{poLCA, Linzer2011} for the latent class model both implemented in R.
In the existing models, the estimation algorithm was iterated until the convergence criterion, $1 \times 10^{-8}$ (tolerance for termination, $\texttt{eps}$ for $\texttt{loglm}$ and $\texttt{tol}$ for $\texttt{poLCA}$), was satisfied.
In the latent class model, random initialization for parameter estimation was repeated 100 times ($\texttt{nrep=100}$), and different numbers of latent classes ($\texttt{nclass}$) were examined.
In the proposed model, the maximum likelihood estimation was performed using the sequential least squares programming (SLSQP) algorithm implemented in $\texttt{scipy.optimize.minimize}$ in Python.
The convergence tolerance of the SLSQP algorithm ($\texttt{ftol}$) was $1 \times 10^{-12}$.

We limited the set of items for analysis to reduce the computational cost.
We used four features ($\texttt{EXT1}$--$\texttt{EXT4}$) on the original five-point scale and five features ($\texttt{EXT1}$--$\texttt{EXT5}$) on a three point scale in comparing the goodness of fit of the proposed model and the log-linear model.
We collapsed the original five-point response scale into a three-point scale by merging responses 1 and 2 into ordinal level 0, retaining response 3 as ordinal level 1, and merging responses 4 and 5 into ordinal level 2.
Table~\ref{table:comparison} summarizes the negative log-likelihood per data point for training and test data $- \bar{l}(\hat{\theta} \mid \mathcal{D})$, and the mean computation time in seconds per single random initialization for parameter estimation, for each model.
Here, the log-likelihood of the training data per sample size at the estimated parameter is defined by 
\begin{align}
    \bar{l}(\hat{\theta} \mid \mathcal{D}_{\mathrm{train}})
    \equiv&
    \frac{1}{N_{\mathrm{train}}} \log p(\mathcal{D}_{\mathrm{train}} \mid \hat{\theta}), \notag \\
    =&
    \frac{1}{N_{\mathrm{train}}} \sum_{i \in \text{train}} \log p(\mathbf{y}_i \mid \hat{\theta}),
    \hspace{0.5cm}
    \hat{\theta} = \mathrm{argmax}_{\theta} \; \bar{l}(\theta \mid \mathcal{D}_{\mathrm{train}}).
\end{align}
In the proposed model, we examined both the direct parametrization of the structure given in Proposition~\ref{proposition:categorical_ordinal} ($\Lambda - I$) and the extended diagonally dominant parametrization ($B C^{-1}$).
From this table, we see that the representational flexibility of the extended diagonally dominant parametrization, which realizes both the nonnegative joint probabilities and the occurrence patterns required for categorical and ordinal variables, is comparable to that of the direct parametrization of the structure given in Proposition~\ref{proposition:categorical_ordinal} that realizes only the required occurrence patterns.
Furthermore, while the proposed model has a reasonable representational flexibility when the number of ordinal levels is small, its flexibility becomes limited compared to the log-linear model when the number of ordinal levels reaches five or more.
As discussed in Section~\ref{subsubsec:mixed_type}, this stems from the fact that the structured parametrization given in Proposition~\ref{proposition:categorical_ordinal} does not have enough parameters to adequately represent joint distributions involving multiple variables with many categories or ordinal levels.
This tendency becomes particularly pronounced as the number of categories or ordinal levels increases.

A common approach to enhance the representational flexibility of a model is to employ a mixture distribution, as in latent class models.
However, employing such latent variable models can also lead to the problem of improper solutions.
In latent class models, a symptom of this problem is that the occurrence probabilities of categorical variables in a specific class take boundary values of 0 or 1 when the number of latent classes increases.
In fact, our numerical results show that, in latent class models, as the number of latent classes increases, the computation time required for parameter estimation increases, and at the same time, the goodness of fit to the validation data gets worse compared to that of the log-linear model.
The limitation on the representational flexibility of the proposed model in the presence of variables with many categories or ordinal levels can be mitigated by using a mixture of Grassmann distributions.
A mixture of the proposed Grassmann distributions may be less susceptible to the problem of improper solutions than a conventional latent class model, because it can potentially achieve the same level of representational flexibility with fewer latent classes than a mixture of independent categorical distributions.

\begin{figure}[htbp]
    \centering
    \includegraphics[scale=1, pagebox=cropbox, viewport= 0 0 296.508 173.033, clip]{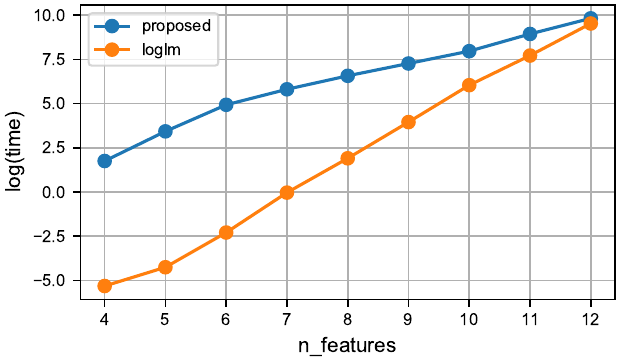}
    \caption{
        Logarithm of the computation time (in seconds) required for parameter estimation in the proposed and log-linear models.
    }
    \label{fig:computational_cost}
\end{figure}

The proposed model exhibits favorable scaling behavior in terms of computational complexity and memory usage compared to the log-linear model.
Fig.~\ref{fig:computational_cost} compares the computation times of the log-linear model and the proposed model as a function of the number of features $f$.
We used up to 12 features ($\texttt{EXT1}$--$\texttt{EST2}$) on the original five-point scale.
For the proposed model, the computation time was averaged over ten optimization runs with different random initializations, all of which satisfied the convergence criterion.
The computation time of the log-linear model increases exponentially with $f$, whereas that of the proposed model exhibits a more gradual increase.
Our empirical results suggest that the computation time of the proposed model scales approximately as $q^3 * \mathrm{n\_unique} * \mathrm{n\_params}^3$, where $\mathrm{n\_unique}$ is the number of unique observed states, and $\mathrm{n\_params}$ is the number of model parameters to be estimated.
Furthermore, while the log-linear model for $f \ge 13$ using the \texttt{loglm} functions cannot fit into memory in our computing environment, the proposed model can still estimate the model parameters in such cases.
However, even in high-dimensional settings where the proposed model offers computational advantages, the number of parameters to be estimated can reach several thousand, and the numerical estimation becomes time-consuming and is difficult to perform on a standard personal computer.
Therefore, to take full advantage of the proposed model, a high-performance computing environment equipped with a graphics processing unit may be preferable.

\begin{table}[hbtp]
    \caption{Goodness of fit and computation time (in seconds) for the proposed model, the log-linear model, and the latent class model.}
    \label{table:comparison}
    \centering
    {
    \renewcommand{\arraystretch}{1.2}
    \begin{tabular}{c|ccc|ccc}
        \hline
         & \multicolumn{3}{c|}{5 features with 3-point scale} & \multicolumn{3}{c}{4 features with 5-point scale} \\
        \hline
        model & $- \bar{l}(\hat{\theta} \mid \mathcal{D}_{\mathrm{train}})$ & $ - \bar{l}(\hat{\theta} \mid \mathcal{D}_{\mathrm{test}})$ & time & $- \bar{l}(\hat{\theta} \mid \mathcal{D}_{\mathrm{train}})$ & $ - \bar{l}(\hat{\theta} \mid \mathcal{D}_{\mathrm{test}})$ & time \\
        \hline
        proposed ($\Lambda - I$) & 4.66694 & 4.65298 & 2.443 & 5.72389 & 5.72212 & 39.141 \\
        proposed ($B C^{-1}$) & 4.66694 & 4.65298 & 3.107 & 5.72389 & 5.72212 & 40.353 \\ 
        $\texttt{loglm}$ & 4.56301 & 4.54639 & 0.005 & 5.51024 & 5.51538 & 0.006 \\
        $\texttt{poLCA(nclass=2)}$ & 4.69571 & 4.68087 & 0.554 & 5.73661 & 5.73321 & 0.547 \\
        $\texttt{poLCA(nclass=4)}$ & 4.58316 & 4.56532 & 4.583 & 5.54480 & 5.54676 & 1.929 \\
        $\texttt{poLCA(nclass=6)}$ & 4.56626 & 4.55116 & 11.250 & 5.51788 & 5.52254 & 8.444 \\
        $\texttt{poLCA(nclass=8)}$ & 4.56093 & 4.54757 & 118.751 & 5.51178 & 5.51833 & 68.846	 \\
        $\texttt{poLCA(nclass=10)}$ & 4.55886 & 4.54690 & 216.994 & 5.50792 & 5.51710 & 119.718 \\
        \hline
    \end{tabular}
    }
\end{table}

\section{Conclusion} \label{sec:conclusion}
We proposed a method for modeling the joint distribution of categorical and ordinal dummy variables using the Grassmann distribution.
Naive applications of the Grassmann distribution to categorical and ordinal dummy variables may lead to numerically pathological solutions in maximum likelihood estimation.
We proposed a specific structure of the parameter matrix of the Grassmann distribution which ensures that undesired co-occurrence probabilities of categorical and ordinal dummy variables are exactly zero.
However, the proposed parameter structure cannot be expressed by the traditional diagonally dominant parametrization that ensures the nonnegativity of the Grassmann probability distribution for binary dummy variables.
Therefore, we proposed a variant of the diagonally dominant parametrization that ensures the nonnegativity of probabilities and is compatible with the proposed parameter structure.
We have theoretically proven that this variant of the diagonally dominant parametrization ensures the nonnegativity of probabilities.
Using artificially generated data, we numerically verified that the proposed distribution reproduces the means of the categorical and ordinal dummy variables successfully and reproduces correlations reasonably well.
We also investigated the sampling distribution of the maximum likelihood estimator using artificially generated data and found numerical evidence suggesting consistency of the estimator.

Although the proposed distribution has a favorable scalability of computational cost and memory usage and can reproduce the correlations among dummy variables for categorical and ordinal variables reasonably well, the representational flexibility is limited compared to that of the existing log-linear model.
Specifically, when there is a strong correlation between variables with many categories or ordinal levels, it becomes difficult for the proposed model to reproduce their correlation.
For this reason, it is preferable for the proposed model to be used as a building block for other statistical models rather than directly modeling the association among categorical and ordinal variables.
Examples of such models include mixture models, as well as state-space models, factor analysis, and partial least squares, for data containing qualitative variables.
We hope that the proposed construction provides a useful basis for developing statistical learning methods for qualitative data.

\section*{Statements and Declarations}
\begin{itemize}
\item Competing Interests and Funding \\
The authors have no competing interests to declare that are relevant to the content of this article. No funding was received for conducting this study.
\item Availability of data and materials \\
The reader data underlying this article are publicly available from the R package \texttt{catdata} (version 1.2.2), which is published under the license of GPL-2~\citep{catdata}.
The Firstborn Personality Scale Responses data underlying this article are publicly available from \href{https://openpsychometrics.org/_rawdata/}{OpenPsychometrics.org}.
\end{itemize}

\begin{appendices}

\section{Inverted Grassmann distribution} \label{sec:inverted_grassmann_distribution}
In this appendix, we provide a brief summary of the Grassmann distribution used in this paper.
The definition of the Grassmann distribution in this paper differs from the Grassmann distribution in the previous studies~\citep{Arai2021, Arai2022} in that the effect of dummy variables is inverted.
This change of definition is intended to make the parameter matrix of the Grassmann distribution nonsingular while realizing the co-occurrence probabilities required for categorical and ordinal dummy variables.
Since the effect of the dummy variables is something like the definition of the sign (plus/minus) of the electric current, such a change is permissible and does not affect the essence of the problem.
We denote the Grassmann distribution used in this paper by $\mathcal{G}^{(1)}$.

\subsection{Grassmann distribution for binary dummy variables}
We denote $q$-dimensional dummy variables by a column vector $\mathbf{y}$, where each element takes a value 0 or 1.
That is, the vector $\mathbf{y}$ is a bit vector with each element taking a value 0 or 1.
We denote the set of all indices of dummy variables as $R \equiv \{1,2,\dots, q\}$.
An index label for dummy variables is divided into two parts with subscripts 1 and 0, for variables that take the value 1 and 0, respectively.
For example, an index label for binary variables $R$ is divided into a subset $R_1 \subseteq R$ and its set difference $R_0 = R \setminus R_1$.
Then, the joint distribution of the Grassmann distribution is expressed as follows:
\begin{align}
    p(\mathbf{y}_{R_1} = \bm{1}, \mathbf{y}_{R_0} = \bm{0}) =& \mathcal{G}^{(1)}(\mathbf{y}_{R_1} = \bm{1}, \mathbf{y}_{R_0} = \bm{0} \mid \Sigma ), \notag \\
    =&
    \det \tilde{\Sigma}, \label{eq:grassmann_joint_Sigma}
\end{align}
where
\begin{align}
    \det \tilde{\Sigma}
    \equiv 
    \det \begin{bmatrix} I - \Sigma_{R_1 R_1} & \Sigma_{R_1 R_0} \\ - \Sigma_{R_0 R_1} & \Sigma_{R_0 R_0} \end{bmatrix},
\end{align}
and the indexed notation $\Sigma_{R_1 R_1}$ expresses the submatrix of $q \times q$ matrix $\Sigma$, with its rows and columns consisting of the index set $R_1$, and $I$ is the identity matrix.
In terms of the inverse matrix $\Lambda \equiv \Sigma^{-1}$, the Grassmann distribution is represented by a principal minor of the matrix $\Lambda - I$:
\begin{align}
    \mathcal{G}^{(1)}(\mathbf{y}_{R_1} = \bm{1}, \mathbf{y}_{R_0} = \bm{0} \mid \Sigma=\Lambda^{-1} )
    = \frac{\det( \Lambda_{R_1 R_1} - I )}{\det \Lambda}.
    \label{eq:grassmann_joint_Lambda}
\end{align}
For all possible joint probabilities to be nonnegative, the matrix $\Lambda - I$ must be a $P_0$-matrix, which means all principal minors of $\Lambda - I$ are nonnegative.
In particular, the diagonal elements of the matrix $\Lambda - I$ must be nonnegative.

To express the marginal and conditional distributions, we first define the notation of index.
We divide the sets of all indices of dummy variables into two subset parts; $R=(S, T)$.
The number of elements in these sets of indices is represented by $q_S$ and $q_T$, these of course satisfy $q_S + q_T = q$.
Then, the subvector of $\mathbf{y}$ comprising the subset of indices $S$ is represented as $\mathbf{y}_S$.
An index label for dummy variables is further divided into two parts with subscripts 1 and 0, for variables that take the value 1 and 0, respectively.
For example, an index label for binary variables $S \subseteq R$ is divided into a subset $S_1 \subseteq S$ and its set difference $S_0 = S \setminus S_1$, where the subvectors $\mathbf{y}_{S_1}$ and $\mathbf{y}_{S_0}$ take the values as $\mathbf{y}_{S_1}=\bm{1}$ and $\mathbf{y}_{S_0}=\bm{0}$, respectively.
Using the index notation described above, the marginal distribution is expressed as
\begin{align}
    p(\mathbf{y}_T) = \mathcal{G}^{(1)}( \mathbf{y}_T \mid \Sigma_{TT}). \label{eq:grassmann_marginal}
\end{align}
The conditional distribution is given by
\begin{align}
    p(\mathbf{y}_S \mid \mathbf{y}_T) =& \mathcal{G}^{(1)}(\mathbf{y}_S \mid \tilde{\Sigma}_{S|T}), \\
    \tilde{\Sigma} =&
    \begin{bmatrix} \tilde{\Sigma}_{SS} & \tilde{\Sigma}_{ST} \\
        \tilde{\Sigma}_{TS} & \tilde{\Sigma}_{TT}
    \end{bmatrix}
    \equiv \begin{bmatrix} \Sigma_{SS} & - \Sigma_{ST_1} & \Sigma_{ST_0} \\
        \Sigma_{T_1 S} & I - \Sigma_{T_1 T_1} & \Sigma_{T_1 T_0} \\
        \Sigma_{T_0 S} & - \Sigma_{T_0 T_1} & \Sigma_{T_0 T_0}
    \end{bmatrix}, \\
    \tilde{\Sigma}_{S|T} \equiv & \tilde{\Sigma}_{SS} - \tilde{\Sigma}_{ST} \tilde{\Sigma}_{TT}^{-1} \tilde{\Sigma}_{TS}, \notag \\
    =& [\Lambda_{SS} - \Lambda_{ST_1} (\Lambda_{T_1 T_1} - I)^{-1} \Lambda_{T_1 S}]^{-1},
\end{align}
where $\Sigma_{TT}^{-1}$ denotes the inverse matrix of the submatrix, $\Sigma_{TT}^{-1} \equiv (\Sigma_{TT})^{-1}$, and $\Sigma_{S|T}$ is the Schur complement.

The mean and covariance of dummy variables are given by
\begin{align}
    E[y_r] =& 1 - \Sigma_{rr}, \\
    \mathrm{Cov}[y_r \, y_s] =& - \Sigma_{rs} \Sigma_{sr}, \hspace{0.5cm} (r \ne s, \hspace{0.3cm} r, \, s \in R).
\end{align}
That is, the diagonal element of $\Sigma$ expresses the marginal probability that the dummy variable takes the value 0. 
Therefore, the diagonal elements of matrices $\Sigma$ and $I - \Sigma$ must be within the closed interval $[0, 1]$.
On the other hand, the conditional mean and covariance with all the other conditioned variables observed to be 0 are given by
\begin{align}
    E[y_r \mid \mathbf{y}_{R \setminus r} = \bm{0}] =&  1 - \frac{1}{\Lambda_{rr}} = \Bigl( \frac{\Lambda_{rr}}{\Lambda_{rr} - 1} \Bigr)^{-1}, \\
    \mathrm{Cov}[y_r \, y_s\mid \mathbf{y}_{R \setminus \{r, s \}}= \bm{0}] =& \frac{-\Lambda_{rs} \Lambda_{sr}}{(\Lambda_{rr} \Lambda_{ss} - \Lambda_{rs}\Lambda_{sr})^2}, \hspace{0.5cm} (r,\,s \in R).
\end{align}
That is, the reciprocal of the diagonal element $\Lambda_{rr}$ expresses the conditional probability that the corresponding dummy variable takes the value 0 with all the other conditioned variables observed to be 0.

Here we should mention the degrees of freedom of the parameters of the Grassmann distribution.
The model parameter $\Sigma$ (respectively $\Lambda$) has redundant degrees of freedom.
Specifically, there are different parameters that generate the exact same joint probability.
In fact, the joint distribution remains unchanged when we multiply the $i$th row of the matrix $\Sigma$ by a constant $c$ while simultaneously multiplying the corresponding $i$th column by $1/c$.
Furthermore, the joint distribution is invariant under matrix transposition.
These degrees of freedom can be represented in matrix form,
\begin{align}
    \Sigma' = D^{-1} \Sigma D, \hspace{0.5cm}\Leftrightarrow&  \hspace{0.5cm}
    \Lambda' = D^{-1} \Lambda D, \label{eq:multiplication_dof} \\
    \Sigma' = \Sigma^T,  \hspace{0.5cm}\Leftrightarrow&  \hspace{0.5cm} \Lambda' = \Lambda^T, \label{eq:transposition_dof}
\end{align}
where $D$ is an arbitrary invertible diagonal matrix.

\subsection{Grassmann distribution for mixed continuous and binary variables}
In this subsection, we summarize the probability distribution that describes the interaction between continuous and binary dummy variables in which the effect of dummy variables is inverted from that of the previous studies~\citep{Arai2021, Arai2022}, to avoid possible confusion.

We denote $p$-dimensional continuous variables and $q$-dimensional dummy variables by column vectors $\mathbf{x}$ and $\mathbf{y}$, respectively.
The dummy vector $\mathbf{y}$ is a bit vector with each element taking a value 0 or 1.
Model parameters of our distribution consist of mean and covariance parameters of a multivariate normal distribution $(\bm{\mu}, \Sigma)$, a $q \times q$ matrix of the Grassmann distribution $\Lambda$~\citep{Arai2021}, and a $q \times p$ matrix $G$ representing the interaction between continuous and binary variables.
The matrix $\Lambda - I$ must be a $P_0$-matrix~\citep{Tsatsomeros2020}, where $I$ is an identity matrix.
Each element of the matrix $G$, $[G]_{sj}, \; (s=1,2,\dots, q, \; \text{and} \; \; j=1,2,\dots,p)$, is also represented by a $p$-dimensional column vector $\mathbf{g}_s$ as
\begin{align}
    G \equiv \begin{bmatrix} \mathbf{g}_1, \;\;  \mathbf{g}_2, \;\; \dots, \;\; \mathbf{g}_q \end{bmatrix}^T
    , \hspace{0.5cm} 
    [G]_{sj} = [\mathbf{g}_{s}^T]_j,
\end{align}
where $T$ stands for matrix transposition.
We denote the set of all indices of continuous and binary variables as $I\equiv \{1,2,\dots,p\}$ and $R \equiv \{1,2,\dots, q\}$, respectively.
An index label for binary variables is divided into two parts with subscripts 1 and 0, for variables that take the value 1 and 0, respectively.
For example, an index label for binary variables $R$ is divided into a subset $R_1 \subseteq R$ and its set difference $R_0 = R \setminus R_1$.
We denote a $q$-dimensional bit vector $\bm{1}_{S}$ for the index label $S \subseteq R$ whose elements take values in $\{0, 1\}$,
\begin{align}
    [\bm{1}_{S}]_s \equiv
    \begin{cases} 1, \hspace{0.5cm} \text{if} \; \; s \in S \\ 
        0, \hspace{0.5cm} \text{if} \; \; s \notin S
    \end{cases}, \; \; s \in R.
\end{align}
Then, the proposed joint distribution is expressed as
\begin{align}
    p(\mathbf{x}, \mathbf{y}=\bm{1}_{R_1}) 
    =&  \pi_{R_1}(\Sigma) \, 
    \mathcal{N}(\mathbf{x} \mid \bm{\mu} + \Sigma G^T \bm{1}_{R_1}, \Sigma), \notag \\
    \equiv & \pi_{R_1}(\Sigma) 
    \frac{1}{(2\pi)^{p/2} \det \Sigma^{1/2}} \, \exp\left\{-\frac{1}{2}(\mathbf{x}- \bm{\mu} - \Sigma G^T \bm{1}_{R_1})^T \Sigma^{-1}(\mathbf{x}- \bm{\mu} - \Sigma G^T \bm{1}_{R_1}) \right\},
    \label{eq:joint} \\
    \pi_{R_1}(\Sigma) \equiv& \frac{\det (\Lambda_{R_1 R_1} -I) \exp\Bigl(\frac{1}{2} \bm{1}_{R_1}^T G \Sigma G^T \bm{1}_{R_1} \Bigr) }{\sum_{R_1' \subseteq R} \det (\Lambda_{R_1' R_1'} -I) \exp\Bigl( \frac{1}{2} \bm{1}_{R_1'}^T G \Sigma G^T \bm{1}_{R_1'} \Bigr) },
\end{align}
where $\Lambda_{R_1 R_1}$ is a submatrix of $\Lambda$, and summation $\sum_{R_1' \subseteq R}$ runs over all possible states of binary variables.
The partition function, the normalization constant, of this distribution is not given analytically, and thus, one has to sum over all possible states of the binary variables to calculate the partition function.
As we will see below, the coefficient $\pi_{R_1}(\Sigma)$ corresponds to the mixing weight of a mixture of Gaussian distributions with equal covariance.
That is, the above joint distribution corresponds to one normal distribution out of $2^q$ mixture of normal distributions.

To express the marginal and conditional distributions, we first define the notation of index.
We denote the index label of a subset of all indices as $J \subseteq I$.
Then, the subvector comprising the subset of indices $J$ is represented as $\mathbf{x}_J$.
We divide the sets of all indices of continuous and binary variables into three subset parts; $I=(J, L, K)$ and $R=(S, U, T)$, where the index labels $L$ and $U$ are introduced to handle missing values.
The number of elements in these sets of indices is represented by $p_J$, $p_L$, $p_K$ and $q_S$, $q_U$, $q_T$, these of course satisfy $p_J + p_L + p_K = p$ and $q_S + q_U + q_T = q$.
Then, the vectors $\mathbf{x}$ and $\mathbf{y}$ can be partitioned into subvectors as $\mathbf{x}=\mathbf{x}_I = [\mathbf{x}_J, \mathbf{x}_L, \mathbf{x}_K]$ and $\mathbf{y}=\mathbf{y}_R=[\mathbf{y}_S, \mathbf{y}_U, \mathbf{y}_T]$, respectively.
Again, an index label for binary variables is further divided into two parts with subscripts 1 and 0, for variables that take the value 1 and 0, respectively.
For example, an index label for binary variables $S \subseteq R$ is divided into a subset $S_1 \subseteq S$ and its set difference $S_0 = S \setminus S_1$, where the subvectors $\mathbf{y}_{S_1}$ and $\mathbf{y}_{S_0}$ take the values as $\mathbf{y}_{S_1}=\bm{1}$ and $\mathbf{y}_{S_0}=\bm{0}$, respectively.
The union of the index label $J$ and $K$ is denoted as $J + K \equiv J \cup K$.
Using the index notation described above, the marginal distribution is expressed as
\begin{align}
    p(\mathbf{x}_K=\mathbf{x}_{I \setminus (J+L)}, \mathbf{y}_T=\mathbf{y}_{R \setminus (S+U)})
    =&
    \sum_{S_1 + U_1 \subseteq R \setminus T} \pi_{R_1}(\Sigma) \, 
    \mathcal{N}(\mathbf{x}_K \mid \bm{\mu}_K + \Sigma_{KI} G^T \bm{1}_{R_1}, \Sigma_{KK}).
    \label{eq:marginal}
\end{align}
In particular, when all binary variables are marginalized, the marginal distribution is precisely a $2^q$ mixture of Gaussian distributions with equal covariance, where mixing weights are given by $\pi_{R_1}(\Sigma)$ and the mean of the normal distributions is shifted by $\Sigma G^T \bm{1}_{R_1}$.
On the other hand, when all continuous variables are marginalized, the marginal distribution is no longer in the same form as the Grassmann distribution.

The conditional distribution with missing values for $\mathbf{x}_L$ and $\mathbf{y}_U$ is given by
\begin{align}
    p(\mathbf{x}_J, \mathbf{y}_S |\mathbf{x}_K, \mathbf{y}_T)
    =& 
    \frac{\sum_{U_1 \subseteq R \setminus (S + T)} \pi_{R_1}\bigl(\Sigma_{(J+L)|K}\bigr)
    \exp \Bigl\{ \bm{1}_{R_1}^T G \Sigma_{IK} \Sigma_{KK}^{-1} (\mathbf{x}_K - \bm{\mu}_K) \Bigr\} }{\sum_{S_1' + U_1' \subseteq R \setminus T} \pi_{R_1'}\bigl(\Sigma_{(J+L)|K}\bigr)
    \exp \Bigl\{ \bm{1}_{R_1'}^T G \Sigma_{IK} \Sigma_{KK}^{-1} (\mathbf{x}_K - \bm{\mu}_K) \Bigr\} } \notag \\
    & \hspace{-2cm}
    \mathcal{N}\bigl(\mathbf{x}_J \mid \bm{\mu}_J + \Sigma_{JK} \Sigma_{KK}^{-1} (\mathbf{x}_K -\bm{\mu}_K) + (\Sigma_{J (J+L)} - \Sigma_{JK} \Sigma_{KK}^{-1} \Sigma_{K(J+L)}) G_{(J+L) R}^T \bm{1}_{R_1}, \Sigma_{J|K} \bigr),
    \label{eq:conditional}
\end{align}
where $\Sigma_{KK}^{-1}$ denotes the inverse matrix of the submatrix $\Sigma_{KK}$, the matrix $\Sigma_{J|K} \equiv \Sigma_{JJ} - \Sigma_{JK}\Sigma_{KK}^{-1}\Sigma_{KJ}$ is the Schur complement, $G_{(J+L)R}^T$ denotes the submatrix of $G^T$ consisting of rows indexed by $(J+L)$ and columns indexed by $R$, and the mixing weight is defined as previously mentioned,
\begin{align}
    \pi_{R_1}\bigl(\Sigma_{(J+L)|K}\bigr) \equiv
    \frac{\det (\Lambda_{R_1 R_1} -I) \exp\Bigl\{ \frac{1}{2} \bm{1}_{R_1}^T G_{R (J+L)} \Sigma_{(J+L)|K} G_{(J+L)R}^T \bm{1}_{R_1} \Bigr\} }{\sum_{R_1' \subseteq R} \det (\Lambda_{R_1' R_1'} -I) \exp \Bigl\{ \frac{1}{2} \bm{1}_{R_1'}^T G_{R (J+L)} \Sigma_{(J+L)|K} G_{(J+L)R}^T \bm{1}_{R_1'} \Bigr\} }.
\end{align}

When there are no missing values, the conditional distribution is expressed more concisely:
\begin{align}
    p(\mathbf{x}_J, \mathbf{y}_S |\mathbf{x}_K=\mathbf{x}_{I \setminus J}, \mathbf{y}_T = \mathbf{y}_{R \setminus S}) 
    =& 
    \frac{ \pi_{R_1}(\Sigma_{J|K}) \exp\Bigl\{ \bm{1}_{S_1}^T G \Sigma_{IK}\Sigma_{KK}^{-1} (\mathbf{x}_K -\bm{\mu}_K) \Bigr\} }{  \sum_{S_1' \subseteq R \setminus T} \pi_{R_1'}(\Sigma_{J|K}) \exp\Bigl\{ \bm{1}_{S_1'}^T G \Sigma_{IK}\Sigma_{KK}^{-1} (\mathbf{x}_K -\bm{\mu}_K) \Bigr\} } \notag \\ 
    & \mathcal{N}\bigl(\mathbf{x}_J \mid \bm{\mu}_J + \Sigma_{JK} \Sigma_{KK}^{-1} (\mathbf{x}_K -\bm{\mu}_K) + \Sigma_{J|K} G_{JR}^T \bm{1}_{R_1}, \Sigma_{J|K} \bigr).
    \label{eq:conditional_distribution_without_marginal_variables}
\end{align}
In particular, when observed variables consist exclusively of continuous variables, the conditional distribution is expressed as a normal distribution,
\begin{align}
    p(\mathbf{x}_J|\mathbf{x}_K=\mathbf{x}_{I \setminus J}, \mathbf{y}_R) =& 
    \mathcal{N} \left( \mathbf{x}_J \mid \bm{\mu}_J + \Sigma_{JK}\Sigma_{KK}^{-1} (\mathbf{x}_K -\bm{\mu}_K) + \Sigma_{J|K} G_{JR}^T \bm{1}_{R_1} , \Sigma_{J|K}\right).
    \label{eq:conditional_distribution_continuous}
\end{align}
Here, the mean of the distribution is shifted depending on the value of the binary variables conditioned.
On the other hand, when observed variables consist exclusively of binary variables, the conditional distribution is expressed as a Grassmann distribution:
\begin{align}
    p(\mathbf{y}_S | \mathbf{x}_I, \mathbf{y}_T=\mathbf{y}_{R\setminus S})
    = &
    \mathcal{G}^{(1)} \bigl(\mathbf{y}_S \mid I + (\Lambda-I)_{S|T_1} \, \mathrm{Exp}\bigl\{ G_{SI} (\mathbf{x}_I - \bm{\mu}_I) \bigr\} \bigr), \notag \\
    \equiv & 
    \frac{ \det \bigl[ (\Lambda-I)_{S_1|T_1} \, \mathrm{Exp} \bigl\{G_{S_1 I} (\mathbf{x}_I - \bm{\mu}_I) \bigr\}  \bigr] }{\det \bigl[ I + (\Lambda - I)_{S|T_1} \, \mathrm{Exp} \bigl\{ G_{SI}(\mathbf{x}_I - \bm{\mu}_I) \bigr\}  \bigr] }, 
    \label{eq:conditional_distribution_binary} \\
    (\Lambda - I)_{S|T_1}  \equiv & \Lambda_{SS}-I - \Lambda_{S T_1}(\Lambda_{T_1 T_1}-I)^{-1}\Lambda_{T_1  S},
\end{align}
where
\begin{align}
    \mathrm{Exp}\bigl\{ G_{SI}(\mathbf{x}_I-\bm{\mu}_I) \bigr\} 
    \equiv 
    E^{G_{SI}(\mathbf{x}_I-\bm{\mu}_I)}
    \equiv 
    \mathrm{diag}\bigl[ \exp \bigl\{\mathbf{g}_s^T (\mathbf{x}_I-\bm{\mu}_I) \bigr\} \bigr], \hspace{0.5cm} s \in S
\end{align}
is a diagonal matrix with nonnegative diagonal elements.

\section{Formulae for a partitioned matrix} \label{sec: matrix_identity}
In this appendix, we describe the formulae of a partitioned matrix in our notation.
For a partitioned matrix of an invertible matrix $\Sigma$ with subsets index $A$ and $B$,
\begin{align}
	\Sigma = \begin{bmatrix} \Sigma_{AA} & \Sigma_{AB} \\ \Sigma_{BA} & \Sigma_{BB} \end{bmatrix}, \hspace{0.5cm}
	\Lambda = \Sigma^{-1} = \begin{bmatrix} \Lambda_{AA} & \Lambda_{AB} \\ \Lambda_{BA} & \Lambda_{BB} \end{bmatrix},
\end{align}
where $\Sigma_{BB}$ is an invertible matrix, the Schur complement of $\Sigma$ with respect to $\Sigma_{BB}$ is defined by
\begin{align}
\Sigma / \Sigma_{BB} \equiv & \Sigma_{A|B}, \notag \\
\equiv &  \Sigma_{AA} - \Sigma_{AB} \Sigma_{BB}^{-1} \Sigma_{BA} .
\end{align}
If the diagonal blocks $\Sigma_{AA}$ and $\Sigma_{BB}$ are both invertible, the inverse of the partitioned matrix is expressed in the following two ways:
\begin{align}
\begin{bmatrix} \Sigma_{AA} & \Sigma_{AB} \\ \Sigma_{BA} & \Sigma_{BB} \end{bmatrix}^{-1}
=& \begin{bmatrix}
\Sigma_{AA}^{-1} + \Sigma_{AA}^{-1} \Sigma_{AB} \Sigma_{B|A}^{-1} \Sigma_{BA} \Sigma_{AA}^{-1} \hspace{0.3cm} & - \Sigma_{AA}^{-1} \Sigma_{AB} \Sigma_{B|A}^{-1} \\
- \Sigma_{B|A}^{-1} \Sigma_{BA} \Sigma_{AA}^{-1} & \Sigma_{B|A}^{-1} 
\end{bmatrix}, \notag  \\
=& \begin{bmatrix}
\Sigma_{A|B}^{-1} & -\Sigma_{A|B}^{-1} \Sigma_{AB} \Sigma_{BB}^{-1} \\
- \Sigma_{BB}^{-1} \Sigma_{BA} \Sigma_{A|B}^{-1} \hspace{0.3cm} & \Sigma_{BB}^{-1} + \Sigma_{BB}^{-1} \Sigma_{BA} \Sigma_{A|B}^{-1} \Sigma_{AB} \Sigma_{BB}^{-1}
\end{bmatrix}
=\begin{bmatrix} \Lambda_{AA} & \Lambda_{AB} \\ \Lambda_{BA} & \Lambda_{BB} \end{bmatrix}, \label{eq:inverse_of_a_partitioned_matrix}
\end{align}
which is known as the partitioned inverse formula, and can be derived using elementary operations for a partitioned matrix.
From the above equations, we can read the Woodbury matrix identity or the matrix inversion lemma,
\begin{align}
	\bigl[ \Sigma_{AA} - \Sigma_{AB} \Sigma_{BB}^{-1} \Sigma_{BA} \bigr]^{-1} =&
 \Sigma_{AA}^{-1} + \Sigma_{AA}^{-1} \Sigma_{AB} \Sigma_{B|A}^{-1} \Sigma_{BA} \Sigma_{AA}^{-1}, \label{eq:Woodbury} \\
	\Sigma_{A \mid B}^{-1} \Sigma_{AB} \Sigma_{BB}^{-1}=& \Sigma_{AA}^{-1} \Sigma_{AB} \Sigma_{B \mid A}^{-1}.
\end{align}
Furthermore, we can read the relation between $\Sigma_{AB}$ and $\Lambda_{AB}$:
\begin{align}
\Sigma_{AB} \Sigma_{BB}^{-1} =& - \Lambda_{AA}^{-1} \Lambda_{AB}, \label{eq: Sigma_AB_left} \\
\Sigma_{AA}^{-1} \Sigma_{AB} =& - \Lambda_{AB} \Lambda_{BB}^{-1}. \label{eq: Sigma_AB_right}
\end{align}
In our notation, the Schur complement formula for the determinant can be written as
\begin{align}
\det \Sigma = \det \Sigma_{AA} \det \Sigma_{B|A} = \det \Sigma_{BB} \det \Sigma_{A|B}. \label{eq: matrix_determinant_lemma}
\end{align}

\end{appendices}

\bibliography{grassmann_v9}

\end{document}